\documentclass[sn-mathphys-num]{sn-jnl}

\usepackage{graphicx}%
\usepackage{multirow}%
\usepackage{amsmath,amssymb,amsfonts}%
\usepackage{amsthm}%
\usepackage{mathrsfs}%
\usepackage[title]{appendix}%
\usepackage[table]{xcolor}%
\usepackage{textcomp}%
\usepackage{manyfoot}%
\usepackage{booktabs}%
\usepackage{adjustbox}
\usepackage{algorithm}%
\usepackage{algorithmicx}%
\usepackage{algpseudocode}%
\usepackage{listings}%
\usepackage{subcaption}
\usepackage{tablefootnote}

\usepackage{tikz}
\usetikzlibrary{trees,positioning,shapes,shadows,arrows.meta}
\usepackage{ifthen}
\usepackage{pgfplots}
\pgfplotsset{compat=1.18} 
\usetikzlibrary{patterns}

\usepackage{color, colortbl}
\definecolor{mygray}{rgb}{0.95,0.95,0.95}
\definecolor{darkgreen}{rgb}{0., 0.67, 0.}
\definecolor{ballblue}{rgb}{0.13, 0.67, 0.8}
\definecolor{darkgrey}{rgb}{0.6, 0.6, 0.6}
\definecolor{magenta}{rgb}{0.79, 0.08, 0.48}
\definecolor{cerulean}{rgb}{0.0, 0.48, 0.65}

\definecolor{blue_blind}{HTML}{648FFF}
\definecolor{violet_blind}{HTML}{785EF0}
\definecolor{magenta_blind}{HTML}{DC267F}
\definecolor{orange_blind}{HTML}{FE6100}
\definecolor{yellow_blind}{HTML}{FFB000}
\definecolor{green_blind}{HTML}{009E73}
\definecolor{red_blind}{HTML}{DC3220}
\definecolor{lightblue_blind}{HTML}{88CCEE}

\def\addlegendimage{\csname pgfplots@addlegendimage\endcsname}

\newenvironment{customlegend}[1][]{%
        \begingroup
        \csname pgfplots@init@cleared@structures\endcsname
        \pgfplotsset{#1}%
    }{%
        \csname pgfplots@createlegend\endcsname
        \endgroup
    }%

\newcommand{\G}{\mathcal{G}}

\newcommand{\V}{\mathcal{V}}
\newcommand{\E}{\mathcal{E}}

\theoremstyle{thmstyleone}%
\theoremstyle{thmstyletwo}%

\theoremstyle{thmstylethree}%

\usepackage{ulem}
\usepackage{xcolor}
\usepackage{lipsum}                     
\usepackage{xargs}                      
\usepackage{multirow}
\usepackage[colorinlistoftodos,prependcaption,textsize=small]{todonotes}
\newcommandx{\ari}[2][1=]{\todo[linecolor=red,backgroundcolor=red!25,bordercolor=red,#1]{AP: #2}}
\newcommandx{\lucio}[2][1=]{\todo[linecolor=blue,backgroundcolor=blue!25,bordercolor=blue,#1]{L: #2}}

\begin{document}

\title[Finding Hidden Swing Voters in the 2022 Italian Elections
Twitter Discourse]{Finding Hidden Swing Voters in the 2022 Italian Elections
Twitter Discourse}

\author[1]{\fnm{Alessia} \sur{Antelmi}}\email{alessia.antelmi@unito.it}
\equalcont{All authors contributed equally to this work.}

\author[2]{\fnm{Lucio} \sur{La Cava}}\email{lucio.lacava@dimes.unical.it}
\equalcont{All authors contributed equally to this work.}

\author[3]{\fnm{Arianna} \sur{Pera}}\email{arpe@itu.dk}
\equalcont{All authors contributed equally to this work.}

\affil[1]{%
    \orgname{University of Turin},%
    \orgaddress{
        \city{Turin},
        \country{Italy}%
    }%
}%

\affil[2]{%
    \orgdiv{DIMES Dept.}, 
    \orgname{University of Calabria},%
    \orgaddress{
        \city{Rende}, 
        \country{Italy}%
    }%
}

\affil[3]{
    \orgname{IT University of Copenhagen},%
    \orgaddress{
        \city{Copenhagen}, 
        \country{Denmark}%
    }%
}

\abstract{The global proliferation of social media platforms has transformed political communication, making the study of online interactions between politicians and voters crucial for understanding contemporary political discourse. In this work, we examine the dynamics of political messaging and voter behavior on Twitter during the 2022 Italian general elections. Specifically, we focus on voters who changed their political preferences over time (swing voters), identifying significant patterns of migration and susceptibility to propaganda messages. Our analysis reveals that during election periods, the popularity of politicians increases, and there is a notable variation in the use of persuasive language techniques, including doubt, loaded language, appeals to values, and slogans. Swing voters are more vulnerable to these propaganda techniques compared to non-swing voters, with differences in vulnerability patterns across various types of political shifts. These findings highlight the nuanced impact of social media on political opinion in Italy.}


\maketitle

\section{Introduction}
With the disruptive diffusion of social media platforms over the past decades, many social interactions have shifted to the online world. The study of online social discourse is nowadays crucial for understanding interaction dynamics that may either impede or enhance processes of opinion formation~\cite{monti2022language}, democratic deliberation~\cite{jennings2021social}, and collective action~\cite{yasseri2016political}. 
In the context of political communication, social media have progressively assumed what was once the role of traditional mass media, serving as the source of information dissemination but also as the point of contact with voters~\cite{stieglitz2013social}. 
Twitter, now X, has been a significant platform for the development of political discourse. Messages are publicly disseminated and easily accessible to all users, fostering interactions that are more open and horizontal compared to other social media platforms commonly used in politics, such as Facebook~\cite{park2013does}. Opinion leaders on Twitter play a pivotal role in encouraging public and political engagement, particularly during elections and other politically significant events, when posts become integral elements of political communication in a hybrid media landscape~\cite{jungherr2016twitter}.
The proliferation of online social media usage, particularly of Twitter, among both politicians and the general public has significantly impacted the study of political discourse, opening new possibilities for investigating the political digital landscape. On the one hand, this shift has paved the way for the exploration of digital communication practices in terms of popularisation, disintermediation, personalisation, intimisation, and populism~\cite{bracciale2017define}. On the other hand, it has enabled scholars to investigate the political stances~\cite{conover2011predicting} and opinions of the general public in real-time~\cite{karami2018mining}, shedding light onto phenomena such as political polarization~\cite{uzogara2023democracy}. Nonetheless, the analysis of political preference shifts through the observation of social media and the relationship between such shifts and the susceptibility to online communication, especially in terms of propaganda, remain to be explored. In the next paragraphs, we will provide some context to these elements and describe existing research that analyses similar matters going beyond the typical US political setting.

\medskip
\noindent\textbf{Related Work.}
The use of persuasion and propaganda messages is a long-established strategy in communication. From the church Reformation in the fifteenth century to the communication of warfare messages during World War I and II, propaganda has historically been employed to focus the attention of masses on certain issues and events, long before the rise of traditional and social media~\cite{cull2003propaganda}. With the broadcasting of political debates on national television, political representatives began harnessing persuasive language to stir public opinion towards their cause~\cite{parry2001rhetorical}.
The disruptive shift towards social media has marked a change in communication style, and politicians have gradually adapted to such a change by exploiting the power of the new media to deliver persuasive messages to their followers in a personalized campaign manner~\cite{enli2013personalized}.
Twitter, in particular, has become a significant tool for politicians during election campaigns. Its use has been shown to positively impact electoral support, with interactive communication enhancing preferential votes~\cite{kruikemeier2014political}. Moreover, on micro-blogging platforms like Twitter, personalized messages foster emotional closeness and positive evaluations of political candidates, especially among individuals with weak party identities~\cite{lee2012personalize}. However, this effect can vary based on the candidates' gender and partisan identity~\cite{mcgregor2018personalization}, stressing the importance of studying voters' vulnerability to such messages. 
A change in political preferences defines what is known as ``swing voter": a voter who is not solidly committed to a single candidate or party over time~\cite{mayer2008exactly}. Politicians specifically target these voters during election campaigns because their likelihood of being persuaded is higher than that of standard voters. Different types of swing voters emerge in election contexts, depending on the depth of commitment to the swing. One example is the so-called ``party switcher"~\cite{mayer2008exactly}: a voter who crosses party lines in subsequent election periods, beyond simply changing the preferred candidate. The presence of various levels of swing and the different vulnerabilities of voters to political messages define an intriguing landscape for election studies.

\smallskip 

The European political landscape generally remains relatively unexplored compared to its US counterpart in the context of online observational studies. Some notable exceptions include research on social media's role in spreading populist messages during the 2018 Andalusian elections~\cite{rivas2020far}, in political marketing during the 2010 UK general elections~\cite{harris2015social}, in terms of extreme political attitudes in the case of the interaction between French Parliament members and regular Twitter users~\cite{peralta2024multidimensional} and in terms of party-based segregation in the Twitter network of Dutch MPs~\cite{tolsma2024twitter}.
In the case of Italy, the volume of online political discourse on social media has recently increased, but the research coverage level does not compare with other Countries such as the US. Nonetheless, researchers focused on the Italian landscape while studying polarization and homophily~\cite{lai2019stance} or the role of populism in online engagement~\cite{bracciale2021does}. 
The Italian political context is characterized by several political parties mainly organized around left and right-wing coalitions, with some space for centrist parties and parties without a clear leaning. Although most alliances remain stable over the years, these coalitions are not static. During election times, and especially with the formation of a new government, parties often regroup into coalitions based on common political programs. These coalitions can include parties from different parts of the political spectrum, as seen in the 65th government of the Italian Republic (2018-2019), where the cabinet was formed by a coalition between Movimento 5 Stelle (an anti-establishment party beyond the left-right distinction) and Lega (a right-wing party). Alternatively, coalitions can be formed by parties with similar values, such as the Terzo Polo coalition created by Azione and Italia Viva during the 2022 general elections. This diverse political landscape necessitates exploring voter dynamics related to the evolving nature of political alliances.
The 2022 Italian elections serve as a particularly interesting case study due to the resignation of Prime Minister Mario Draghi and the rising popularity of the populist right. This election marks a significant shift in Italy's political landscape, highlighting the dynamics and implications of populist movements gaining traction in Europe.

\medskip
\noindent \textbf{Contribution.\ } In this study, we aim to fill the existing research gap concerning the analysis of political preference shifts in online observational studies and the patterns of susceptibility to propaganda within the Italian social media political landscape. We achieve this goal by exploring the interplay between political messaging on Twitter and the behavior of ``swing voters'' in the context of the 2022 Italian general elections. In particular, we pose the following research questions (RQs):

\begin{description}
    \item[{RQ$_1$}.] How can the Twitter discourse surrounding the 2022 Italian general elections be modeled as a network? What are its key structural traits?

    \smallskip
    
    \item [{RQ$_2$}.] Who are the most popular politicians? What primary propaganda techniques were exploited by political representatives during the 2022 elections?

    \smallskip
    
    \item[{RQ$_3$}.] Can we identify ``swing voters'' within the Italian political online sphere during the elections? Are there variations in susceptibility to political propaganda techniques between swing voters and non-swing voters?

\end{description}

Our findings reveal a stable political discourse in Italy, characterized by homogeneous political communities. While users generally tend to remain within the same or similar (in terms of political leaning) communities over time, there are significant patterns of migration that highlight the presence of swing voters. During election periods, the popularity of political representatives exhibits a dynamic trend, typically increasing during the electoral campaign and diminishing post-election. Politicians play different roles within the social network of political debate: some maintain a ``core" role throughout the election period, while others gain prominence only afterward. By analyzing the language used by political representatives, we identified \textit{doubt}, \textit{loaded language}, \textit{appeal to values}, and \textit{slogans} as the most frequently employed propaganda techniques during election times with notable differences in usage across political wings. A variation in the use of such techniques by political representatives is observed together with a change in their popularity levels. Overall, swing voters appear more vulnerable to propaganda compared to non-swing voters, though the specific vulnerability patterns vary among different types of swing voters.

\medskip
\noindent \textbf{Article organization.\ }
The remainder of this work is organized as follows. Section~\ref{sec:methodology} describes the data and methodology we used for answering the listed research questions. Section~\ref{sec:results} contains the network-based and language-based analyses conducted, while Section~\ref{sec:discussion} discusses the results of these analyses. 
Finally, Section~\ref{sec:conclusions} concludes our work and provides pointers for future discussions and research directions.

\section{Methodology}\label{sec:methodology}

\begin{figure}[h!]
    \centering
    \includegraphics[width=\linewidth]{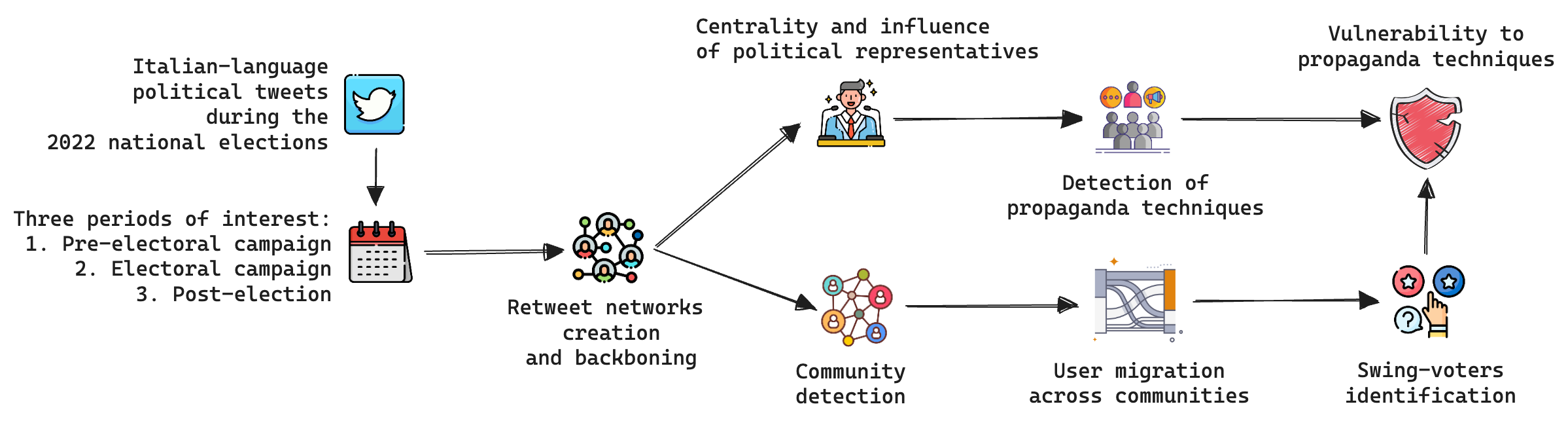}
    \caption{Workflow of the analyses followed in this work.}
    \label{fig:workflow}
\end{figure}

This section describes the data we used in our work and the methodology we followed to answer the RQs outlined in the Introduction. Figure~\ref{fig:workflow} visually summarizes the steps of our analyses.

\subsection{Data}
We performed our analyses by leveraging the first publicly available dataset of Italian-language political conversations during the 2022 Italian general elections~\cite{Pierri_CIKM_2023}. The dataset spans a period of four months, from July to October 2022, covering the most significant events of the Italian political landscape during that period. It contains 19,087,594 tweets shared by 618,089 unique users. 

This dataset also includes the social media handles of political representatives, which are mapped to their corresponding main political parties. These include left-wing parties, such as \textit{Alleanza Verdi e Sinistra} (AVS), \textit{Partito Democratico}~(PD), centrist parties, like \textit{Movimento 5s} (M5s), \textit{Azione - Italia Viva} (Az-Iv) and \textit{Noi Moderati} (NM), and right-wing parties, such as \textit{Forza Italia} (FI), \textit{Fratelli d'Italia} (FdI), and \textit{Lega} (L). 
Table~\ref{tab:vol_tweets_party} provides an overview of the data, while Table~\ref{tab:party_leader} lists each party along with its acronym, political spectrum, and leader.

\begin{table}[t!]
\centering
\footnotesize

\caption{Main statistics of the Twitter data used in this work.}
\label{tab:vol_tweets_party}


\begin{tabular}{l|l|rrr}

     \toprule
     
     \textbf{Split} &  
     \textbf{Party} & 
     \textbf{\# Users} & 
     \textbf{\# (Tweets + Quotes)} & 
     \textbf{\# Retweets} 
     \\

     \midrule
     All users &  
     NA & 
     618,089 & 
     8,654,779 & 
     10,639,195\\
     
     \midrule
     
     Non-Politicians & 
     NA & 
     617,508 & 
     8,636,771 & 
     10,623,875\\
     
     \midrule
     
     \multirow{8}{*}{Politicians} & 
     
       Alleanza Verdi e Sinistra & 12 & 622 & 1,979 \\
     & Azione - Italia Viva  & 30 & 3,570 & 3,176 \\
     & Forza Italia    & 62 & 1,207 & 363 \\
     & Fratelli d'Italia    & 181   & 4,371 & 3,350 \\
     & Lega            & 95        & 3,869 & 2,575 \\
     & Movimento 5s       & 80     & 1,039 & 413 \\
     & Noi moderati        & 15    & 161 & 39 \\
     & Partito Democratico  & 107  & 3,169   & 3,425 \\   
     \bottomrule
\end{tabular}

\end{table}
\begin{table}[b!]
\centering
\footnotesize

\caption{Association between political party, leader, and political spectrum. \label{tab:party_leader}}

\begin{tabular}{lrll}

     \toprule
      
     \textbf{Party} & 
     \textbf{Acronym} &
     \textbf{Political spectrum} & 
     \textbf{Leader} 
     \\

     \midrule
     
     Alleanza Verdi e Sinistra &
     AVS &
     Left wing &
     Angelo Bonelli, Nicola Fratoianni 
     \\

     Azione - Italia Viva  &
     Az-Iv &
     Terzo Polo & 
     Carlo Calenda 
     \\

     Forza Italia & 
     FI &
     Right wing & 
     Silvio Berlusconi 
     \\

     Fratelli d'Italia & 
     FdI &
     Right wing & 
     Giorgia Meloni 
     \\

     Lega & 
     Lega &
     Right wing & 
     Matteo Salvini 
     \\
     
     Movimento 5s & 
     M5s &
     Other & 
     Giuseppe Conte 
     \\
     
     Noi moderati & 
     NM &
     Right wing & 
     Maurizio Lupi
     \\
     
     Partito Democratico & 
     PD &
     Left wing & 
     Enrico Letta 
     \\ 
     
     \bottomrule
\end{tabular}

\end{table}

\subsection{Network Modeling and Structural Analysis}\label{subsec:network_methodology} 

This section details how we defined our network model on the 2022 Italian elections Twitter discourse and presents the set of analyses we carried out to answer \textbf{RQ$_1$}, and the first part of \textbf{RQ$_2$} regarding the centrality of politicians.
We point out that, throughout our work, we decided to use retweets as a proxy for political endorsement and common interest, as they have been proven effective in previous works~\cite{Gamout2018frenchelections,Metaxas_ICWSM_2021}.

\subsubsection{Network Definition}
\label{subsubsec:net_def}

Let $\V$ be the set of users \textit{tweeting} or \textit{retweeting} in our dataset. We can define a directed weighted network $G = \langle \V, \E, w \rangle$, where $\E \subseteq \V \times \V$ denotes the set of \textit{retweet} relations, such that the edge $(u, v) \in \E$ with $u, v \in \V$ indicates that the user $u$ retweeted the user $v$, and $w: \E \mapsto \mathbb{R}$ is a weighing function such that $w(u, v)$ stores the number of times $u$ retweeted $v$. 

In addition, to also analyze the evolution of the political landscape on Twitter, we inferred three sub-networks representing \textit{(i)} the period before the election campaign (01 July - 25 August 2022), referred to as \textit{before} or \textit{pre-electoral campaign}; \textit{(ii)} the official electoral campaign period (26 August - 24 September 2022), referred to as \textit{during} or \textit{electoral campaign}; and \textit{(iii)} the election days and subsequent period (25 September - 31 October 2022), referred to as \textit{after} or \textit{post-election}.
Formally, given an interval $t \in$\{\textit{before}, \textit{during}, \textit{after}\}, we can infer the corresponding induced directed weighted sub-networks as $G_t = \langle \V_t, \E_t, w_t \rangle$, where $\V_t \subseteq \mathcal{V}$ models the set of users active during $t$, $\E_t \subseteq \E$ encompasses their retweet interactions during $t$, and $w_t: \E_t \mapsto \mathbb{R}$ behaves like $w$ but only considering the interval $t$. We indicate these networks as $G_b = \langle \V_b, \E_b, w_b \rangle$, $G_c = \langle \V_c, \E_c, w_c \rangle$, and $G_a = \langle \V_a, \E_a, w_a \rangle$.

\paragraph{Network backboning}
To reduce the noise associated with online social ties and only considering the true significant structure of the observed interactions, we pruned the three retweet networks $G_b$, $G_c$, and $G_a$ by removing spurious edges or ties formed due to random chance via a backboning approach.

\smallskip

In this work, we used a combination of statistical backbone extraction techniques. Statistical methods assess the significance of edges in a network by employing hypothesis testing techniques based on either empirical distribution or a predefined null model. These methods calculate p-values for each edge and filter out edges based on their p-values. Specifically, we used the following backbone extraction methods implemented in the Python library \texttt{netbone}~\cite{Yassin_SciRep_2023}.

\begin{itemize}
    \item \textit{Disparity Filter} \cite{Serrano_PNAS_2009}.
        This is one of the first methods developed for network backboning. The disparity filter evaluates the strength and degree of each node locally. The null hypothesis is that the strength of a node is distributed uniformly at random over the node's incident edges.

    \item \textit{Marginal Likelihood Filter} \cite{Dianati_PhysRevE_2016}.
        While the Disparity Filter evaluates the significance of each edge independently based on the nodes it connects, the Marginal Likelihood filter considers the two nodes connected by the edge. It treats an integer-weighted link as comprising multiple unit edges. The null model assumes that each unit edge randomly selects two nodes, resulting in a binomial distribution. In essence, it computes the probability of drawing at least ``w" unit edges from the network's strength (the sum of all weights), with the probability proportional to the strengths of both nodes.

    \item \textit{Noise Corrected Filter} \cite{Coscia_ICDE_2017}.
        Similar to the Marginal Likelihood Filter, this method assumes that edge weights are drawn from a binomial distribution. However, it estimates the probability of observing a weight connecting two nodes within a Bayesian framework. This approach facilitates the generation of posterior variances for all edges, which, in turn, allows for the creation of confidence intervals for each edge weight. Edges are then removed if their weight deviates from the expectation by a certain number of standard deviations (the sole parameter of the algorithm).

    \item \textit{Locally Adaptive Network Sparsification Filter} \cite{Foti_PlosOne_2011}.
        This method makes no assumptions about the underlying weight distribution, but it uses the empirical cumulative density function to evaluate the statistical significance of the edges. In a nutshell, it calculates the probability of choosing an edge randomly with a weight equal to the observed weight.
        
\end{itemize}
After applying all four extraction techniques, we pruned edges based on a significance level of 0.05. Finally, we derived the consensus backbone by intersecting all four backbones. We independently repeated this process on each temporal network and obtained the three consensual backbones $G'_b$, $ G'_c$, and $ G'_a$ associated with $G_b$, $G_c$, and $G_a$, respectively.

\subsubsection{Network Analysis}\label{subsubsec:net_analysis}
In the following, we describe the algorithms and metrics we used to analyze the community composition of the backbone networks and the structural role played by the politicians, especially by party leaders.

\paragraph{Community Detection, Analysis, and Dynamics}

\textit{Community detection.}
We resorted to the well-known Louvain community detection algorithm~\cite{Blondel2008louvain} to discover communities in our political retweet networks, using its weighted implementation. This algorithm is particularly suited for large-scale networks and operates in an unsupervised manner, meaning it does not require the number of communities to be specified in advance.
Louvain is conceived as a two-step hierarchical greedy optimization method to maximize modularity $M$. Specifically, the first step involves (i) \textit{modularity optimization}, and the second step involves (ii) \textit{community aggregation}. These steps are iteratively performed until no further improvements in modularity can be achieved.

\smallskip

Formally speaking, the modularity M of a network partition is defined as:
\begin{equation}
    M = \frac{1}{2m} \sum_{i,j} \left[ A_{ij} - \frac{k_i k_j}{2m} \right]\delta(c_i, c_j),
\end{equation}
where $A_{ij}$ is the entry in the binary adjacency matrix between nodes $i$ and $j$, $k_i$ and $k_j$ denote the degrees of nodes $i$ and $j$, respectively, $m$ indicates the total number of edges in the network, and the Kronecker delta function $\delta(c_i, c_j)$ outputs 1 if the two nodes $i$ and $j$ pertain to the same community, 0 otherwise.

\smallskip

Initially, each node is assigned to its own singleton community. In the first phase (i.e., {modularity optimization}), the Louvain algorithm seeks to achieve the maximum positive gain in modularity by moving each node $i$ from its current community to a different community $C$. The gain in modularity, $\delta M$, is computed as follows:
\begin{equation}
    \Delta M = \left[ \frac{\sum_{in} + 2k_{i,in}}{2m} - \left( \frac{\sum_{tot} + k_i}{2m} \right)^2 \right] - \left[ \frac{\sum_{in}}{2m} - \left( \frac{\sum_{tot}}{2m} \right)^2 - \left( \frac{k_i}{2m} \right)^2 \right]
\end{equation}
where $\sum_{\text{in}}$ is the sum of links within the target community $C$, $\sum_{\text{tot}}$ is the sum of links incident to nodes in $C$, $k_i$ is the sum of links incident to node $i$, $k_{i,\text{in}}$ is the sum of links from node $i$ to nodes in $C$, and $m$ is the total number of edges in the network. This phase continues until no more improvements in modularity can be achieved.

\smallskip

In the second phase (i.e., {community aggregation}), the algorithm merges each of the communities identified in the previous step into a single giant node to reduce the network's complexity. The first phase is then reapplied to this newly formed network. Both phases—modularity optimization and community aggregation—are iteratively performed until no further increase in modularity is possible.

\medskip
\noindent \textit{Community analysis.}
After computing the community structure of the three backbones $G'_b$, $G'_c$, and $G'_a$, we evaluated their composition in terms of political parties. Specifically, for each community, we examined whether it contained a politician. If a community did contain a politician, we labeled that community with the name of the politician's party. In cases where a community contained two or more representatives from different parties, we assigned all the parties found as labels to that community.

\medskip
\noindent \textit{Community dynamics.}
To assess the stability and persistence of the communities found in each retweet network, we build two migration matrices describing how many users transitioned from one community to another between consecutive timeframes. 
In more detail, for each community $c$ found in the network $G'_b$, we evaluated the percentage of users transitioning from $c$ to another community $c'$ found in the network $ G'_d$. This percentage was normalized by the size of the origin community $c$. We followed the same procedure for assessing the user flow from communities in $ G'_d$ to $ G'_a$.
Based on the labels assigned to each political and non-political community in the previous step, we manually aligned them over time to examine potential migration processes among community members.

\paragraph{Popularity Metrics}
To assess the popularity of politicians within the retweet networks, we used the following metrics.
\begin{itemize}
    \item \textit{In-strength.}
        Given the underlying nature of the political networks we are analyzing, we used this metric as a proxy for user engagement toward the political representatives. Hence, the higher the in-strength, the more engagement and attention the politician has received from other users. 

        \smallskip
        
    \item \textit{PageRank.}
        We leveraged this metric as a measure of politician centrality within the retweet networks. Thus, the higher the ranking the more important the politician is from a structural point of view.

        \smallskip

    \item \textit{VoteRank.} 
        Similar to PageRank, the VoteRank algorithm~\cite{Zhang_SciRep_2016} computes a ranking of nodes in a network based on incoming links. In this algorithm, each node votes for its in-neighbors, and the node receiving the highest number of votes is elected iteratively. In subsequent turns, the voting ability of the out-neighbors of elected nodes is decreased.
        We applied the VoteRank algorithm to our political networks because it operates on the principle that if person A supports person B, the support strength of A to others generally diminishes. Hence, this approach would allow us to identify the candidates with the strongest support within the network.
\end{itemize}

\paragraph{K-Core Decomposition}
To better identify the significance of users within our backbones, we decomposed these networks into groups of nodes based on their relevance. To achieve this, we used \textit{core decomposition}, a process that assigns each node in the network an integer called the \textit{core number}, which indicates how well the node is connected with respect to its neighbors. This efficient process, which operates linearly with respect to the number of edges in the network, is particularly suited for detecting the local importance of nodes~\cite{malliaros2020core}. Core decomposition determines a hierarchical decomposition of the network into nested subnetworks, using a threshold (k) based on properties of the network~\cite{SeidmanCore}. In our setting, we considered the in-degree of a node, as it represents the number of retweets a given user receives, serving as a proxy for its relevance within the network.

\smallskip

Formally, given a network $\G$ and for any choice of an integer value $k \geq 0$, we can say that a subnetwork $\G_k \subseteq \G$ is the \textit{k-core} of $\G$ if it is a maximal subnetwork of $\G$ in which all nodes have degree at least $k$. Moreover, we denote with the \textit{degeneracy} of $\G$ the maximum value $k$ such that $\G_k \neq \emptyset$, and the corresponding network $\G_k$ is dubbed \textit{inner-most core}. The set of all obtainable $k$-cores $\G_0 \supseteq \G_1 \supseteq ... \supseteq \G_k$ determines the \textit{core decomposition} of the network.

\smallskip

In our scenario, by performing the core decomposition of our retweet networks for each time window in the 2022 Italian general elections, we aim to decompose the political debate into groups of relevance. This approach would allow us to highlight users living in the ``core'' of the debate, and users left in the ``periphery.''

\subsection{Swing Voters Detection and Vulnerability to Propaganda Techniques} \label{subsec:swing_voters_method}
This section delineates the process for extracting propaganda techniques from politicians' messages to answer \textbf{RQ$_2$}, describes how we detected swing voters, and defines the vulnerability of users to such techniques, contributing to \textbf{RQ$_3$}.

\paragraph{Swing Voters Detection}
With the term \textit{swing voter}, we refer to a user who changes their supporting party over a given observation period~\cite{mayer2008exactly}. To capture the emergence of the swing voter phenomenon, we leveraged a mesoscopic analysis at the community level (described in $\S$Community Detection, Analysis, and Dynamics of Section~\ref{subsubsec:net_analysis}) to conduct a microscopic analysis at the user level. In this case, we associated each user in each timeframe with the political label of the community they belonged to. Essentially, we analyzed the migration process of users who switched their community of reference between subsequent election periods. 
We define several ways in which this political shift, or swing, can occur, allowing us to identify and understand the behavior of swing voters in our dataset.

\begin{itemize}
    \item \textit{Hard swing voter}. 
        A user who transitions from one political community to another, where the new community does not include any of the parties from their previous community nor any party within the same political coalition (e.g., from \textit{PD, AVS} to \textit{FdI}).

        \smallskip
    
    \item \textit{Soft swing voter}. 
        A user who shifts from one political community to another, where the new community contains parties that belong to the same coalition as some of the parties in the previous community (e.g., from \textit{PD, M5s} to \textit{AVS, NM}).
        
        \smallskip
    
    \item \textit{Spurious swing voter}. 
        A user who switches from one political community to another where the new community contains parties that were also present in the previous community (e.g., from \textit{PD, AVS, NM} to \textit{PD}).

    \item \textit{Apolitical-to-political swing voter} or \textit{vice-versa}. 
        A user who shifts either from a community that is not labeled as political (i.e., not involving any political representatives) to another that is political (e.g., from \textit{apolitical} to \textit{FdI}), or vice versa (e.g., from \textit{PD, NM} to \textit{apolitical}).

\end{itemize}
All users not involved in any of the previously mentioned types of voting swings are labeled as \textit{non-swing voters}.

\paragraph{Propaganda Detection}
The use of propaganda, that is the dissemination of information used to influence public opinion, has been extensive in politics.
Propaganda detection has been widely explored in the context of online social media, being referred to as \textit{computational propaganda}~\cite{dasan2021survey}. Different tools have been proposed for such a kind of task, ranging from text-based analysis~\cite{rashkin2017truth, barron2019proppy, da2019fine} to network approaches~\cite{liu2017holoscope, chetan2019corerank, pacheco2020unveiling}. 
In particular, Da San Martino et al.~\cite{da2019fine} proposed a method to perform fine-grained propaganda detection of text, allowing for the identification of propaganda techniques described at a textuafragment levelel. The list of 18 propaganda techniques derives from previous literature and comprises the following: \textit{loaded language, name-calling or labeling, repetition, exaggeration or minimization, doubt, appeal to fear/prejudice, flag-waiving, causal oversimplification, slogans, appeal to authority, black-and-white fallacy/dictatorship, thought-termination clichè, whataboutism, reductio ad Hitlerum, red herring, bandwagon, obfuscation/intentional vagueness/confusion} and \textit{straw man}. We refer the reader to the original paper for an in-depth description and theoretical reference of each technique. 

\smallskip

In this work, we adapted the methodology proposed by Da San Martino et al.~\cite{da2019fine} and considered the proposed list of techniques to extract propaganda messages from the tweets posted by political representatives during the 2022 Italian Elections. Formally, given a tweet $t$ composed of overlapping textual fragments $f_1, f_2, ..., f_i, ..., f_n$, and the trained propaganda detection model $m$, we applied $m(t)$ to extract the propaganda techniques used in each fragment $f_i$, if any.

\paragraph{Vulnerability to Propaganda}
We define a user's vulnerability to a propaganda technique as their act of endorsing the given technique at a specific moment. On Twitter, this endorsement is manifested through the act of retweeting a tweet that explicitly references the propaganda technique. Therefore, for every tweet posted by a political representative during a defined election period that includes references to one or more propaganda techniques, we systematically identify the users who retweeted the tweet. In this work, this action indicates the susceptibility of these users to the identified propaganda technique(s).

\smallskip

Formally, let $T_{rep}$ denote the set of tweets posted by political representatives during the designated election period. Each tweet $t$ in $T_{rep}$ contains textual content that may (or may not) reference one or more propaganda techniques. For a given tweet $t$, let $U_t$ represent the set of users who retweeted it. Thus, if $t$ contains any reference to propaganda techniques, the set of users vulnerable to the propaganda expressed in tweet $t$ is formally defined as the whole set $U_t$.

\smallskip

To perform a comparative study of propaganda vulnerability and determine whether swing voters are more susceptible to certain propaganda techniques than non-swing voters, we retrieved a subset of non-swing voters as a baseline. Specifically, we selected this subset of users through the following process. For each period of interest, we identified the set of users who were not labeled as swing voters during that period. We then performed a stratified sampling procedure based on network communities to obtain a subset of non-swing voters with a volume comparable to that of swing voters during the reference period.

\section{Results}
\label{sec:results}
In this section, we describe the key structural characteristics of the retweet networks and discuss their community composition (see Section~\ref{subsec:network_results}). In the second part, we detail the role of the most popular political representatives and focus on analyzing the use of propaganda techniques in the Twitter discourse on Italian elections (see Section~\ref{subsec:politician_influence}). Finally, in the third part, we investigate the existence of swing voters and the impact of propaganda techniques on them (see Section~\ref{subsec:swing_voters_results}).

\subsection{Network Modeling and Structural Analysis}
\label{subsec:network_results} 

In the following, we detail the structural traits of the three retweet networks analyzed, examine their community composition, and observe how these communities change over the three periods of interest.

\subsubsection{Network Definition}
\label{subsubsec:net_definition_results}

Table~\ref{tab:net_stats} lists the structural traits of the retweet networks analyzed, reporting for each period the statistics for the original network and the associated backbone obtained as detailed in Section~\ref{subsubsec:net_def}.

\begin{table}[t!]

\caption{Main structural statistics of the retweet networks analyzed, split by observation period.}
\label{tab:net_stats}

{
    \footnotesize
    \setlength{\tabcolsep}{3pt}
    
    \begin{tabular}{lrrrrrr}
        \toprule
    
         & 
         \multicolumn{2}{c}{\textbf{Pre campaign}} & \multicolumn{2}{c}{\textbf{Electoral campaign}} & \multicolumn{2}{c}{\textbf{Post elections}} \\
         
        \textbf{Feature} & 
        \multicolumn{1}{c}{$G_b$} &  
        \multicolumn{1}{c}{$G'_b$} & 
        \multicolumn{1}{c}{$G_c$} &  
        \multicolumn{1}{c}{$G'_c$} & 
        \multicolumn{1}{c}{$G_a$} &  
        \multicolumn{1}{c}{$G'_a$} \\

        \midrule
        
        \#nodes & 
        120,614 & 
        12,008 & 
        183,688 & 
        18,150 & 
        253,854 & 
        15,943 \\

        \rowcolor{gray!10}
        \#edges & 
        1,009,968 & 
        38,209 & 
        1,366,268 & 
        56,953 & 
        1,325,370 & 
        42,864 \\
        
        Reciprocity & 
        0.0 & 
        0.0 & 
        0.0 & 
        0.0 & 
        0.0 & 
        0.0 \\

        \rowcolor{gray!10}
        Density & 
        6.9$\mathrm{e}^{-5}$  & 
        2.65$\mathrm{e}^{-4}$ & 
        4.0$\mathrm{e}^{-5}$  & 
        1.72$\mathrm{e}^{-4}$ & 
        2.06$\mathrm{e}^{-4}$ & 
        1.69$\mathrm{e}^{-4}$ \\ 
        
        Avg. degree* & 
        16.747 &
        6.363  & 
        14.876 & 
        6.276 & 
        10.442 & 
        5.377 \\

        \rowcolor{gray!10}
        Avg. in-degree & 
        8.373 & 
        3.181 & 
        7.438 & 
        3.138 & 
        5.221 & 
        2.689 \\
        

        
        Degree assortativity* & 
        -0.261 & 
        -0.353 & 
        -0.180 & 
        -0.233 & 
        -0.055 & 
        -0.195 \\

        \rowcolor{gray!10}
        Degree assortativity (out, in) & 
        -0.193 & 
        -0.357 & 
        -0.174 & 
        -0.255 & 
        -0.145 & 
        -0.244 \\
        
        Transitivity* & 
        0.0203 & 
        0.0054 & 
        0.0248 & 
        0.0065 & 
        0.0057 & 
        0.0041 \\

        \rowcolor{gray!10}
        Clustering coefficient* & 
        0.0703 & 
        0.0333 & 
        0.0662 & 
        0.0467 & 
        0.0411 & 
        0.0248 \\


        \rowcolor{gray!10}
        \# weakly connected components & 
        200 & 
        39 & 
        1730 & 
        377 & 
        2507 & 
        382 \\

        Avg. shortest path length** & 
        3.834* & 
        0.112 & 
        4.106* & 
        0.130 & 
        4.130* & 
        0.086 \\

        \rowcolor{gray!10}
        Diameter** & 
        13* & 
        11* & 
        20* & 
        13* & 
        15* & 
        15* \\

        \bottomrule

        \multicolumn{7}{l}{*Statistic calculated by discarding edge orientation.} \\
        \multicolumn{7}{l}{**Statistic evaluated on the largest connected components.} \\
    \end{tabular}
}

\end{table}

\begin{figure}[b!]
    \begin{subfigure}{\columnwidth}
        \centering
        \begin{tikzpicture}
            \begin{customlegend}[
                legend columns=2,
                legend cell align={left},
                legend style={
                    font=\small,
                    align=left,
                    draw=none,
                    column sep=2ex,
                },
                legend entries={
                    Original networks,
                    Backbone networks
                }
            ]
                \addlegendimage{blue, mark=*, line width=1.15pt}
                \addlegendimage{red, mark=square*, line width=1.15pt}
            \end{customlegend}
        \end{tikzpicture}
    \end{subfigure}
    
    \begin{subfigure}{0.28\columnwidth}
        \begin{tikzpicture}
    \begin{axis}[
        width=1.3\columnwidth, 
        grid=major,
        grid style={dashed,gray!30},
        title = \textit{Pre-campaign},
        xlabel= In-degree,
        ylabel = {CCDF}, 
        ylabel near ticks,
        xlabel near ticks, 
        ylabel style = {
            font=\small
        },
        xlabel style = {
            font=\small
        },
        yticklabel style = {
            font=\footnotesize,
            xshift=0.5ex
        },
        xticklabel style = {
            font=\footnotesize,
            yshift=0.5ex
        },
        title style = {
            font=\small,
            yshift=-0.5ex
        },
        legend pos=south east,
        legend cell align={left},
        xmode=log,
        ymode=log
    ]
    
        \addplot
        table[x=DEG,y=CS,col sep=comma] {./images/networks/data/ccdf_before.txt};
        
        \addplot
        table[x=DEG,y=CS,col sep=comma, mark=square,
        samples=20] {./images/networks/data/ccdf_before_backbone.txt};
        
    \end{axis}
\end{tikzpicture}
    \end{subfigure}
    ~
    \begin{subfigure}{0.28\columnwidth}
        \hspace*{-0.4cm}
        \begin{tikzpicture}
    \begin{axis}[
        width=1.3\columnwidth, 
        grid=major,
        grid style={dashed,gray!30},
        title = \textit{Electoral campaign},
        xlabel= In-degree,
        xlabel near ticks,
        ylabel = CCDF, 
        ylabel style={
            opacity=0
        },
        xlabel style = {
            font=\small
        },
         yticklabel style = {
            font=\footnotesize,
            xshift=0.5ex
        },
        xticklabel style = {
            font=\footnotesize,
            yshift=0.5ex
        },
        title style = {
            font=\small,
            yshift=-0.5ex
        },
        legend pos=south east,
        legend cell align={left},
        xmode=log,
        ymode=log
    ]
    
        \addplot
        table[x=DEG,y=CS,col sep=comma] {./images/networks/data/ccdf_during.txt};
        
        \addplot
        table[x=DEG,y=CS,col sep=comma, mark=square,
        samples=20] {./images/networks/data/ccdf_during_backbone.txt};
        
    \end{axis}
\end{tikzpicture}
    \end{subfigure}
    ~
    \begin{subfigure}{0.28\columnwidth}
        \hspace*{-0.4cm}
        \begin{tikzpicture}
    \begin{axis}[
        width=1.3\columnwidth, 
        grid=major,
        grid style={dashed,gray!30},
        title = \textit{Post-election},
        xlabel= In-degree,
        xlabel near ticks,
        ylabel = CCDF, 
        ylabel style={
            opacity=0
        },
        xlabel style = {
            font=\small
        },
         yticklabel style = {
            font=\footnotesize,
            xshift=0.5ex
        },
        xticklabel style = {
            font=\footnotesize,
            yshift=0.5ex
        },
        title style = {
            font=\small,
            yshift=-0.5ex
        },
        legend pos=south east,
        legend cell align={left},
        xmode=log,
        ymode=log
    ]
    
        \addplot
        table[x=DEG,y=CS,col sep=comma] {./images/networks/data/ccdf_after.txt};
        
        \addplot
        table[x=DEG,y=CS,col sep=comma, mark=square] {./images/networks/data/ccdf_after_backbone.txt};
        
    \end{axis}
\end{tikzpicture}
    \end{subfigure}

    \caption{Comparison of the Complementary Cumulative Distribution Function (CCDF) for the in-degree distributions across the six retweet networks, split by observation period.}
    \label{fig:ccdf_indegree}
\end{figure}
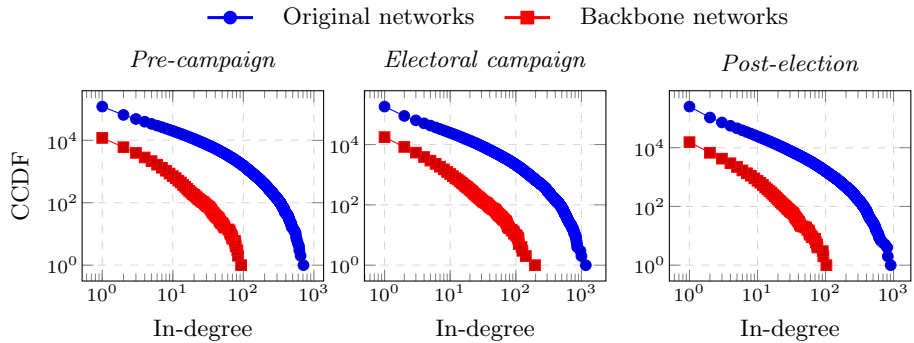

\begin{itemize}
    \item Throughout all three periods, the backbone networks consistently preserve between 6.28\% and 9.96\% of nodes and 3.23\% to 4.16\% of edges. Focusing exclusively on these backbone networks, we observe that the number of the most critical nodes and edges remains stable over the whole observation period. In contrast, the original networks exhibit variability in these numbers across the three timeframes, with a notably higher number of nodes in the period following the elections.

    \smallskip

    \item All networks exhibit a reciprocity value of zero, which is unsurprising given the nature of the network being modeled. In this context, users typically retweet the tweets of political representatives, who rarely retweet other users. 
    The in-degree distribution, shown in Figure~\ref{fig:ccdf_indegree}, further confirms this phenomenon. This distribution is heavily skewed, indicating that a small fraction of users receive the majority of retweets. 
    Additionally, all networks exhibit a negative degree assortativity, once again suggesting that highly retweeted users tend not to connect with one other.

    \smallskip

    \item The backbones exhibit more connected structures than the original networks, as evidenced by the reduced number of weakly connected components, shorter average shortest path length, and smaller diameter. However, all networks present consistently low density, transitivity, and clustering coefficient values. As just discussed, this lack of cohesion was expected, given the nature of the underlying system.
\end{itemize}

\subsubsection{Community Detection, Analysis, and Dynamics}\label{subsubsec:community_analysis}

Table~\ref{tab:comm_stats} reports the main structural traits of the communities identified by the Louvain algorithm in the three backbone networks. These inferred networks are highly clustered, as indicated by consistently robust modularity and coverage metrics, along with negligible conductance values. Due to the high number of communities found, particularly in the second and third periods, we focused on communities with more than 20 users. These larger communities encompass 98.63\%, 93.62\%, and 93.29\% of all nodes in the respective periods.

\begin{table}[b!]
\centering

\caption{Main structural traits of the retweeting communities. \label{tab:comm_stats}}

{\footnotesize

    \setlength{\tabcolsep}{3pt}
    \begin{tabular}{rcccccc}
        \toprule
    
        \multicolumn{1}{c}{\textbf{Period}} &  
        \textbf{\#Comms} & 
        \textbf{
            \begin{tabular}[c]{@{}c@{}}\#Comms with \\ $>$ 20 users \end{tabular}
        } & 
        \textbf{
            \begin{tabular}[c]{@{}c@{}}\# Political \\ comms \end{tabular}
        } & 
        \textbf{Modularity} & 
        \textbf{Coverage} & 
        \textbf{Conductance}  \\
    
        \midrule
        
        Pre-campaign & 
        57 & 
        11 &
        6 &
        0.67 &
        0.86 &
        0.010 
        \\

        Campaign & 
        438 &  
        14 &
        5 &
        0.71 &
        0.89 &
        0.001
        \\

        Post-election & 
        428 &
        16 &
        9 &
        0.70 &
        0.87 &
        0.001
        \\
        \bottomrule
    \end{tabular}
}
\end{table}




    
    
        



\smallskip

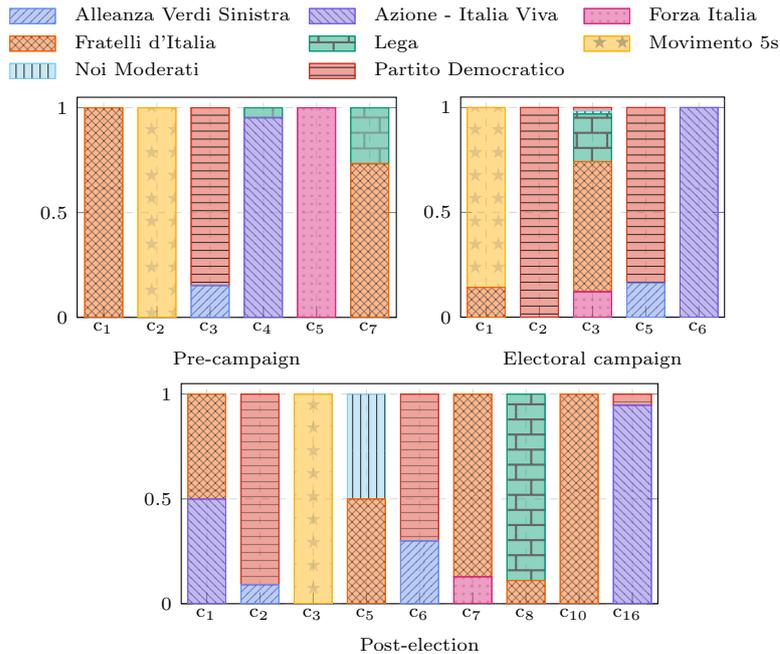
\begin{figure}[t!]
    \begin{subfigure}{\columnwidth}
        \centering
        \begin{tikzpicture}
            \begin{customlegend}[
                legend columns=3,
                legend style={
                    font=\footnotesize,
                    align=left,
                    draw=none,
                    column sep=5pt,
                },
                legend entries={
                    Alleanza Verdi Sinistra,
                    Azione - Italia Viva,
                    Forza Italia,
                    Fratelli d'Italia,
                    Lega,
                    Movimento 5s,
                    Noi Moderati,
                    Partito Democratico
                },
                legend cell align={left},
                legend image code/.code={%
                    \filldraw[
                        #1, 
                        fill opacity = 0.45
                    ] (0cm,-0.1cm) rectangle (0.6cm,0.1cm);
                },
            ]
            \addlegendimage{
                blue_blind, 
                postaction={
                    pattern=north east lines,
                    pattern color = black!80
                }
            }
            \addlegendimage{
                violet_blind, 
                postaction={
                    pattern=north west lines,
                    pattern color = black!80
                }
            }
            \addlegendimage{
                magenta_blind, 
                postaction={
                    pattern=dots,
                    pattern color = black!80
                }
            }
            \addlegendimage{
                orange_blind, 
                postaction={
                    pattern=crosshatch,
                    pattern color = black!80
                }
            }
            \addlegendimage{
                green_blind, 
                postaction={
                    pattern=bricks,
                    pattern color = black!60
                }
            }
            \addlegendimage{
                yellow_blind, 
                postaction={
                    pattern=fivepointed stars,
                    pattern color = black!60
                }
            }
            \addlegendimage{
                lightblue_blind, 
                postaction={
                    pattern=vertical lines,
                    pattern color = black!80
                }
            }
            \addlegendimage{
                red_blind, 
                postaction={
                    pattern=horizontal lines,
                    pattern color = black!80
                }
            }
            \end{customlegend}
        \end{tikzpicture}
    \end{subfigure}
    
    \begin{subfigure}{0.5\columnwidth}
        \flushright









\usetikzlibrary{patterns}

\begin{tikzpicture}


    \begin{axis}[
        ybar stacked,
        x=0.7cm, 
        bar width=0.5cm, 
        height=4.5cm,
        grid=major,
        grid style={dashed,gray!30},
        xlabel= {\footnotesize Pre-campaign},
        xlabel style={
        },
        symbolic x coords={
            1,
            2,
            3,
            4,
            5,
            7
        },
        xtick = data,
        xticklabels = {
            c$_1$,
            c$_2$,
            c$_3$,
            c$_4$,
            c$_5$,
            c$_7$,
         },
        xticklabel style =              
        {
            anchor=east,
            rotate=0,
            font=\footnotesize,
            xshift=7pt,
            yshift=-4pt
        }, 
        xtick pos = bottom,
        yticklabel style = {font=\footnotesize},
        cycle list={
            {blue_blind},
            {violet_blind},
            {magenta_blind},
            {orange_blind},
            {green_blind},
            {yellow_blind},
            {lightblue_blind},
            {red_blind}
        },
        ymin=0, 
        ymax=1.05,
        enlarge x limits=0.1,
        every axis plot/.append style={fill,fill opacity=0.45},
        ]

        %
        %
        \addplot+[
            postaction={
                pattern=north east lines,
                pattern color = black!80
            }
        ]
        table[meta=comm_id,y=AlleanzaVerdiSinistra,col sep=comma] {./images/communities/before_community_comp_data.csv};

        %
        %
        \addplot+[
            postaction={
                pattern=north west lines,
                pattern color = black!80
            }
        ]
        table[meta=comm_id,y=Azione-ItaliaViva,col sep=comma] {./images/communities/before_community_comp_data.csv};

        %
        %
        \addplot+[
            postaction={
                pattern=dots,
                pattern color = black!80
            }
        ]
        table[meta=comm_id,y=ForzaItalia,col sep=comma] {./images/communities/before_community_comp_data.csv};
        	
        %
        %
        \addplot+[
            postaction={
                pattern=crosshatch,
                pattern color = black!80
            }
        ]
        table[meta=comm_id,y=FratellidItalia,col sep=comma] {./images/communities/before_community_comp_data.csv};

        %
        %
        \addplot+[
            postaction={
                pattern=bricks,
                pattern color = black!60
            }
        ]
        table[meta=comm_id,y=Lega,col sep=comma] {./images/communities/before_community_comp_data.csv};

        %
        %
        \addplot+[
            postaction={
                pattern=fivepointed stars,
                pattern color = black!60
            }
        ]
        table[meta=comm_id,y=Movimento5s,col sep=comma] {./images/communities/before_community_comp_data.csv};

        %
        %
        \addplot+[
            postaction={
                pattern=vertical lines,
                pattern color = black!80
            }
        ]
        table[meta=comm_id,y=NoiModerati,col sep=comma] {./images/communities/before_community_comp_data.csv};

        %
        %
        \addplot+[
            postaction={
                pattern=horizontal lines,
                pattern color = black!80
            }
        ]
        table[meta=comm_id,y=PartitoDemocratico,col sep=comma] {./images/communities/before_community_comp_data.csv};
	
    \end{axis}
\end{tikzpicture}
	
    \end{subfigure}
    ~
    \begin{subfigure}{0.5\columnwidth}
        \flushleft









\usetikzlibrary{patterns}

\begin{tikzpicture}


    \begin{axis}[
        ybar stacked,
        x=0.7cm, 
        bar width=0.5cm, 
        height=4.5cm,
        grid=major,
        grid style={dashed,gray!30},
        xlabel= {\footnotesize Electoral campaign},
        xlabel style={
        },
        symbolic x coords={
            1,
            2,
            3,
            5,
            6
        },
        xtick = data,
        xticklabels = {
            c$_1$,
            c$_2$,
            c$_3$,
            c$_5$,
            c$_6$,
         },
        xticklabel style =              
        {
            anchor=east,
            rotate=0,
            font=\footnotesize,
            xshift=7pt,
            yshift=-4pt
        }, 
        xtick pos = bottom,
        yticklabel style = {
            font=\footnotesize,
        },
        ylabel style={
            align=center,
            font=\small,
            yshift=-.4cm
        },
        cycle list={
            {blue_blind},
            {violet_blind},
            {magenta_blind},
            {orange_blind},
            {green_blind},
            {yellow_blind},
            {lightblue_blind},
            {red_blind}
        },
        ymin=0, 
        ymax=1.05, 
        enlarge x limits=0.12,
        every axis plot/.append style={fill,fill opacity=0.45},
        ]

        %
        %
        \addplot+[
            postaction={
                pattern=north east lines,
                pattern color = black!80
            }
        ]
        table[meta=comm_id,y=AlleanzaVerdiSinistra,col sep=comma] {./images/communities/during_community_comp_data.csv};

        %
        %
        \addplot+[
            postaction={
                pattern=north west lines,
                pattern color = black!80
            }
        ]
        table[meta=comm_id,y=Azione-ItaliaViva,col sep=comma] {./images/communities/during_community_comp_data.csv};

        %
        %
        \addplot+[
            postaction={
                pattern=dots,
                pattern color = black!80
            }
        ]
        table[meta=comm_id,y=ForzaItalia,col sep=comma] {./images/communities/during_community_comp_data.csv};
        	
        %
        %
        \addplot+[
            postaction={
                pattern=crosshatch,
                pattern color = black!80
            }
        ]
        table[meta=comm_id,y=FratellidItalia,col sep=comma] {./images/communities/during_community_comp_data.csv};

        %
        %
        \addplot+[
            postaction={
                pattern=bricks,
                pattern color = black!60
            }
        ]
        table[meta=comm_id,y=Lega,col sep=comma] {./images/communities/during_community_comp_data.csv};

        %
        %
        \addplot+[
            postaction={
                pattern=fivepointed stars,
                pattern color = black!60
            }
        ]
        table[meta=comm_id,y=Movimento5s,col sep=comma] {./images/communities/during_community_comp_data.csv};

        %
        %
        \addplot+[
            postaction={
                pattern=vertical lines,
                pattern color = black!80
            }
        ]
        table[meta=comm_id,y=NoiModerati,col sep=comma] {./images/communities/during_community_comp_data.csv};

        %
        %
        \addplot+[
            postaction={
                pattern=horizontal lines,
                pattern color = black!80
            }
        ]
        table[meta=comm_id,y=PartitoDemocratico,col sep=comma] {./images/communities/during_community_comp_data.csv};
	
    \end{axis}
\end{tikzpicture}
	
    \end{subfigure}
     
    \begin{subfigure}{\columnwidth}
        \centering









\usetikzlibrary{patterns}

\begin{tikzpicture}


    \begin{axis}[
        ybar stacked,
        x=0.7cm, 
        bar width=0.5cm, 
        height=4.5cm,
        grid=major,
        grid style={dashed,gray!30},
        xlabel= {\footnotesize Post-election},
        xlabel style={
        },
        symbolic x coords={
            1,
            2,
            3,
            5,
            6,
            7,
            8,
            10,
            16
        },
        xtick = data,
        xticklabels = {
            c$_1$,
            c$_2$,
            c$_3$,
            c$_5$,
            c$_6$,
            c$_7$,
            c$_8$,
            c$_{10}$,
            c$_{16}$,
         },
        xticklabel style =              
        {
            anchor=east,
            rotate=0,
            font=\footnotesize,
            xshift=7pt,
            yshift=-4pt
        }, 
        xtick pos = bottom,
        yticklabel style = {
            font=\footnotesize,
        },
        ylabel style={
            align=center,
            font=\small,
            yshift=-.4cm
        },
        cycle list={
            {blue_blind},
            {violet_blind},
            {magenta_blind},
            {orange_blind},
            {green_blind},
            {yellow_blind},
            {lightblue_blind},
            {red_blind}
        },
        ymin=0, 
        ymax=1.05, 
        enlarge x limits=0.06,
        every axis plot/.append style={fill,fill opacity=0.45},
        ]

        %
        %
        \addplot+[
            postaction={
                pattern=north east lines,
                pattern color = black!80
            }
        ]
        table[meta=comm_id,y=AlleanzaVerdiSinistra,col sep=comma] {./images/communities/after_community_comp_data.csv};

        %
        %
        \addplot+[
            postaction={
                pattern=north west lines,
                pattern color = black!80
            }
        ]
        table[meta=comm_id,y=Azione-ItaliaViva,col sep=comma] {./images/communities/after_community_comp_data.csv};

        %
        %
        \addplot+[
            postaction={
                pattern=dots,
                pattern color = black!80
            }
        ]
        table[meta=comm_id,y=ForzaItalia,col sep=comma] {./images/communities/after_community_comp_data.csv};
        	
        %
        %
        \addplot+[
            postaction={
                pattern=crosshatch,
                pattern color = black!80
            }
        ]
        table[meta=comm_id,y=FratellidItalia,col sep=comma] {./images/communities/after_community_comp_data.csv};

        %
        %
        \addplot+[
            postaction={
                pattern=bricks,
                pattern color = black!60
            }
        ]
        table[meta=comm_id,y=Lega,col sep=comma] {./images/communities/after_community_comp_data.csv};

        %
        %
        \addplot+[
            postaction={
                pattern=fivepointed stars,
                pattern color = black!60
            }
        ]
        table[meta=comm_id,y=Movimento5s,col sep=comma] {./images/communities/after_community_comp_data.csv};

        %
        %
        \addplot+[
            postaction={
                pattern=vertical lines,
                pattern color = black!80
            }
        ]
        table[meta=comm_id,y=NoiModerati,col sep=comma] {./images/communities/after_community_comp_data.csv};

        %
        %
        \addplot+[
            postaction={
                pattern=horizontal lines,
                pattern color = black!80
            }
        ]
        table[meta=comm_id,y=PartitoDemocratico,col sep=comma] {./images/communities/after_community_comp_data.csv};
	
    \end{axis}
\end{tikzpicture}
	
    \end{subfigure}

    \caption{Party composition of the retweeting communities (with size $\geq20$). Party labels are assigned proportionally to the political representatives in each community. Communities with no political representatives are omitted.}
    \label{fig:comms_composition}

\end{figure}

Fig.~\ref{fig:comms_composition} illustrates the political parties associated with each community, with party labels assigned proportionally to the political representatives within them. Interestingly, we found that only 6 out of 11, 5 out of 14, and 9 out of 16 communities (in the three periods, respectively) were labeled as political, indicating that the remaining communities did not include any political representatives.
Further investigation into the composition of these communities revealed their highly homogeneous nature in terms of political affiliation. In other words, most communities were primarily composed of members of a single political party or cohesive coalition. Notably, the distribution of political representatives within these communities reflects what was the real-world political landscape of the Italian elections.
In more detail, we can note almost a one-to-one correspondence between each community and a specific political affiliation in the first observation period. In the case of two political affiliations, both groups belong to the same political wing (left-wing for community c$_3$ and right-wing for community c$_7$).
During the second period, which coincides with the electoral campaign, we observe how also the online platform reveals the rise of the newly born right-wing coalition~(c$_3$), including FdI (led by Giorgia Meloni), Lega (led by Matteo Salvini), FI (led by Silvio Berlusconi), and NM (led by Maurizio Lupi). In contrast, the left-wing coalition appears more fragmented, with one community comprising only representatives from PD (c$_2$) and another community including both PD and AVS (c$_5$). Additionally, Movimento 5 Stelle (led by Giuseppe Conte) and Azione - Italia Viva (led by Carlo Calenda) chose to run for the elections independently, without joining any coalition (c$_1$ and c$_6$, respectively).
In the final period, post-election, we observe a more fragmented situation, where almost every community can again be mostly associated with a single party. However, unlike the first period, we can note multiple communities associated with the same party. For example, communities c$_1$, c$_5$, c$_7$, and c$_{10}$ are all linked to FdI.

\smallskip

\begin{figure}[t!]
    \centering 
    
    \begin{subfigure}{.72\columnwidth}
        \centering
        \includegraphics[width=\textwidth]{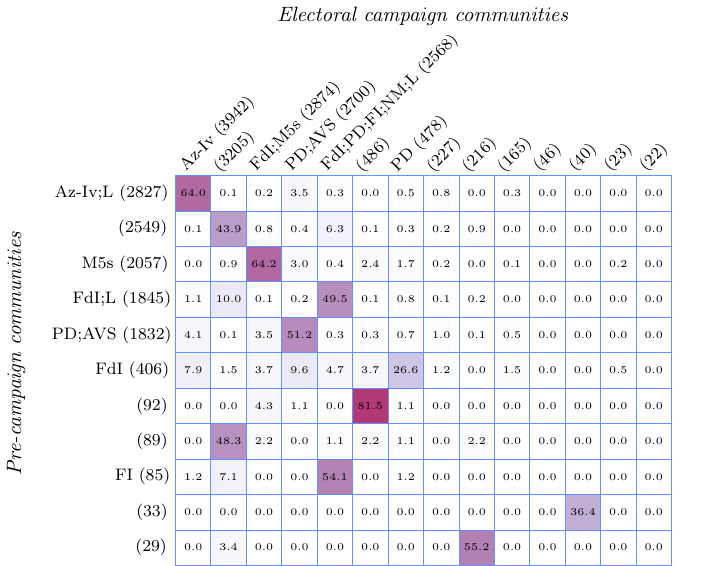}
    \end{subfigure}

    \vspace{1cm}
    
    \begin{subfigure}{.72\columnwidth}
        \includegraphics[width=\textwidth]{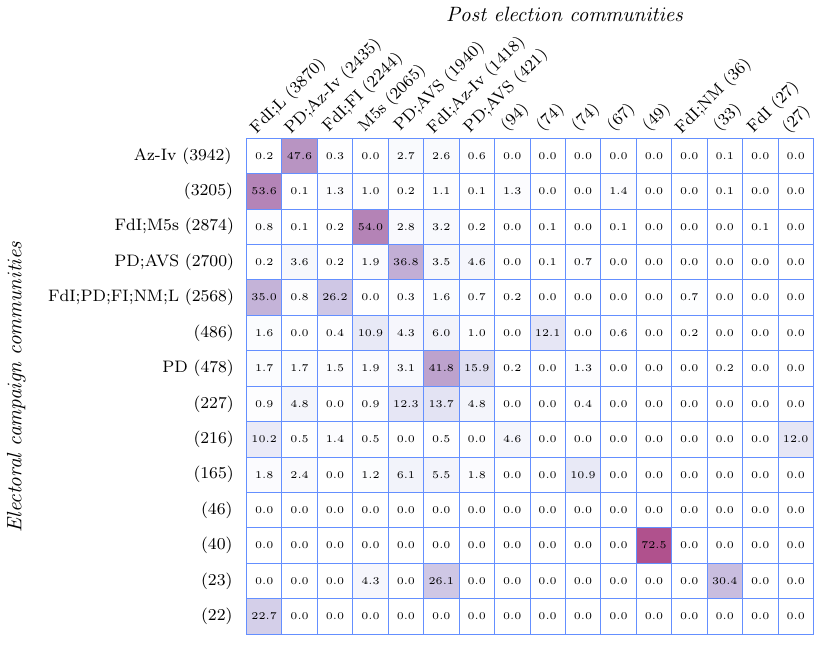}
    \end{subfigure}
    
    \caption{User migration across communities between consecutive timeframes, with numbers in parentheses indicating the size of each community. Political communities are labeled with the parties included within them. The missing percentage of users primarily results from their inactivity in the second and third periods, as the considered communities include most users.}

    \label{fig:migration_matrix}
\end{figure}

%
%
After analyzing the static composition of each community, we extracted the migration matrices illustrating how many users transitioned from one community to another between consecutive timeframes (see Fig.~\ref{fig:migration_matrix}). In this case, our aim was to assess the persistence of communities over the three periods of interest and to identify potential swing voters.
Upon examining these migration matrices, we found limited evidence of political swing voters as most users either remained within their original communities or moved to other communities aligned with their existing political affiliations.
An exception to this trend was observed in 26.6\% of users (i.e., 108 users) who transitioned from a right-wing community (FdI) to a left-wing one (PD) from the pre-election to the electoral campaign period, and 41.8\% of users (i.e., 200 users) who moved from PD to the FdI;Az-Iv group post-election. Another notable pattern is the migration of users from non-political communities to political ones and vice versa, particularly during the transition from the electoral campaign to the post-election period. These migrations included non-political users moving to right-wing (e.g., FdI and Lega, FdI and Az-Iv), left-wing (PD and AVS), and M5S communities. More detailed results on the identification of swing voters are discussed in Section~\ref{subsubsec:swing_voters_detection_results}.
Overall, these findings suggest a relatively stable political discourse on Twitter.

\subsection{Political Representatives and Propaganda} \label{subsec:politician_influence}
In this section, we describe the roles played by the most prominent political representatives in the three retweet networks in terms of centrality measures and k-core decomposition. We then focus on the propaganda techniques used by political representatives on Twitter during the election times.

%
%
\subsubsection{Popularity of Political Representatives} %
\label{subsubsec:centrality}

\begin{table}[b!]


\caption{Summary of centrality statistics based on the in-strengths of the political representatives.}
\label{tab:centralities}

{
    \footnotesize
    \setlength{\tabcolsep}{3pt}
    
    \begin{tabular}{lrrr}
        \toprule
    
         & \textbf{Pre-campaign} & \textbf{Electoral campaign} & \textbf{Post-election} \\

        \midrule

        Max In-Strength & 
        1,547 & 
        3,388 & 
        1,134 \\ 

        \rowcolor{gray!10}
        Avg. In-Strength & 
        110.1 & 
        177.893 & 
        79.419 \\ 

        Stdev. In-Strength & 
        239.002 & 
        446.265 & 
        170.874 \\ 

        \rowcolor{gray!10}
        Median In-Strength & 
        23.0 & 
        31.0 & 
        17.0 \\
        
        \bottomrule

    \end{tabular}
}

\end{table}

Table~\ref{tab:centralities} provides an overview of the main popularity statistics for politicians across the three key periods. Specifically, since politicians' engagement benefits not only from their \textit{reach} (the number of users interacting with their posts) but also from the frequency of these interactions, we resorted to the \textit{in-strength} distribution, a metric accounting for edge weight, thereby considering the frequency of interactions, while computing in-degrees.
As shown in the table, the maximum in-strength value more than doubles from the pre-campaign period to the electoral campaign then decreases to an even lower value after the elections. The average in-strength follows a similar trend across these three periods. These results suggest that while the electoral campaign effectively mobilized attention, there is still room for improving post-election engagement. 
Additionally, the standard deviations and median value indicate that this engagement was not evenly distributed among all politicians but concentrated on a select few, highlighting the presence of more impactful politicians and stressing the need for improved engagement strategies for others.

\smallskip

Interestingly, politicians who managed to engage their audiences before the campaign maintained their engagement levels during the electoral campaign ($r=0.87$, $p<0.001$) and after the elections ($r=0.76$, $p<0.001$). Furthermore, the peaks in engagement achieved during the campaign had notable effects on the engagement levels post-election ($r=0.86$, $p<0.001$).
A detailed analysis of these trends revealed that Luigi Marattin (Az-Iv) experienced the highest increase in engagement from the period before the campaign to the electoral campaign phases. Carlo Calenda (Az-Iv) showed the most significant gain in engagement transitioning from the campaign to the post-election phase, while Raffaella Paita (Az-Iv) saw the largest gain in engagement from the start of the campaign to the post-election phase. Notably, all these politicians belong to the same political party. This consistent rise in engagement suggests that Az-Iv's approach during these phases was particularly effective and warrants a closer examination of their social campaign strategies.
Conversely, Alberto Bagnai (Lega) experienced the largest decrease in absolute engagement from the period before the campaign to the campaign phase, while Giuseppe Conte (M5s) saw the most significant decline in engagement from the start of the elections to the post-election phase. Interestingly, Luigi Marattin (Az-Iv), despite his earlier surge, experienced the most notable drop in engagement from the campaign to the post-election phase. This pattern suggests a transient surge in engagement during the early stages of the elections, highlighting the need for further investigations to understand the underlying causes of such sharp declines. 
Additional insights into plausible explanations for such observed trends are reported in Section~\ref{subsec:propaganda-usage}.

\smallskip

We extended our popularity analysis to include PageRank scores to better capture the centrality of political representatives, and the results are summarized as follows. Unlike the trends observed in engagement, the three election phases significantly influenced the centrality of politicians. Although many politicians maintained their central positions from the pre-campaign phase to the campaign phase ($r=0.73$, $p<0.001$), notable shifts in centrality occurred when moving from the campaign to the post-election phase ($r=0.52$, $p<0.001$). Even greater changes were observed when comparing the start of the elections with the post-election period ($r=0.48$, $p<0.001$).
Our analysis revealed that Luigi Marattin (Az-Iv) gained the most centrality in the early stages of the elections, while Matteo Renzi (Az-Iv) experienced the greatest increase in centrality in the late stages. Notably, Renzi also achieved the highest overall gain in centrality throughout the election period. Conversely, Alberto Bagnai (Lega) lost the most centrality in the early stage, while Elena Bonetti (Az-Iv) and Carlo Cottarelli (PD) experienced the largest drops in centrality in the late stage and throughout the election period, respectively.

\smallskip

Finally, we complemented our centrality analysis of the representatives with VoteRank. In this context, Carlo Calenda (Az-Iv), Giuseppe Conte (M5s), and Giorgia Meloni (FdI) consistently ranked among the top three politicians. In particular, Calenda led the rankings in the first two phases of the elections, while Meloni took the lead in the post-election phase. The rankings showed partial consistency across all election phases (before-during: $r=0.62$, $p<0.001$; during-after: $r=0.64$, $p<0.001$; before-after: $r=0.60$, $p<0.001$).

%
%
\subsubsection{K-Core Decomposition} 

\begin{figure*}[t!]
    \centering
    \setlength{\tabcolsep}{0pt}
    \begin{tabular}{ccc}
         \includegraphics[width=0.33\textwidth]{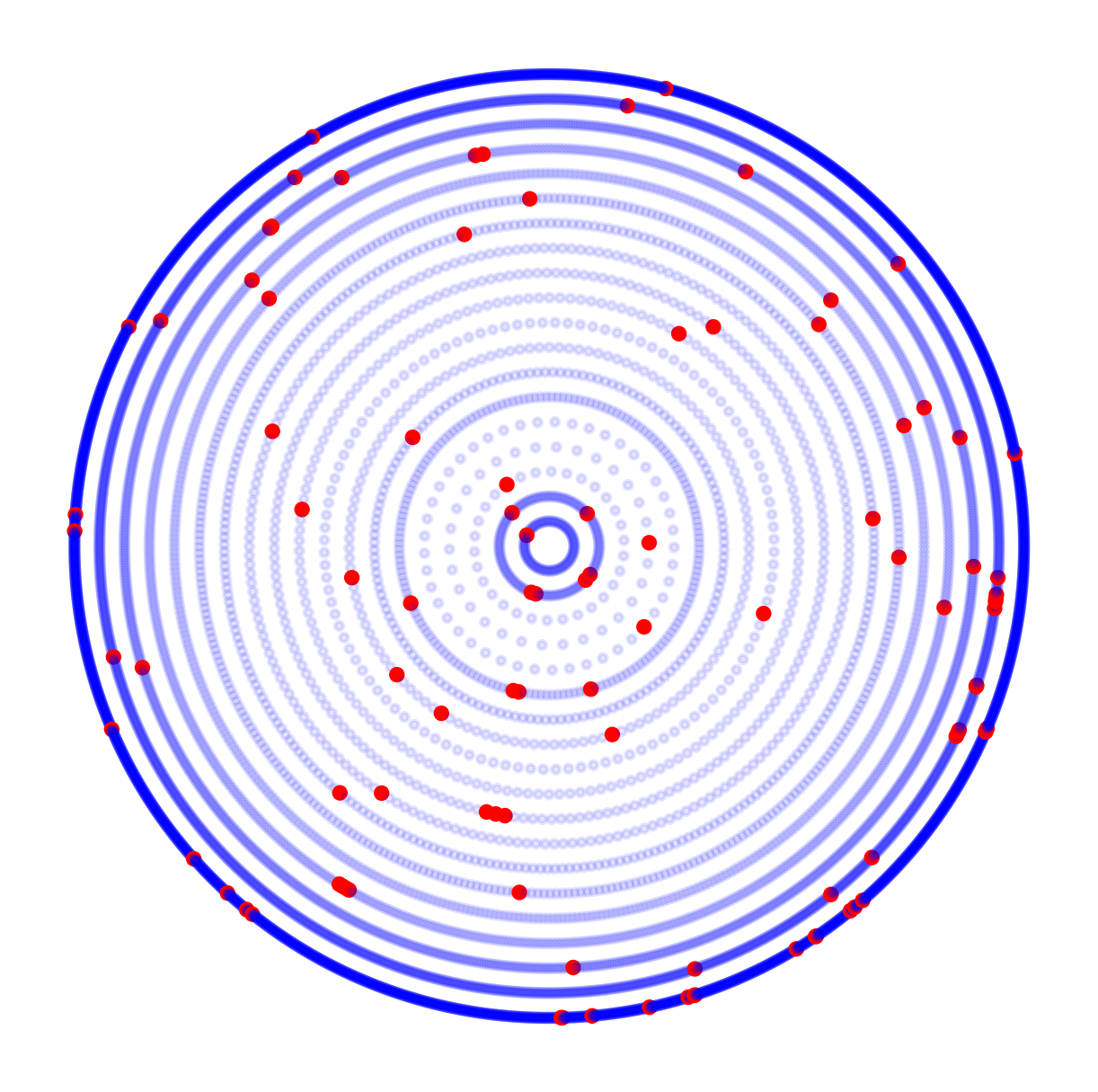} &
         \includegraphics[width=0.33\textwidth]{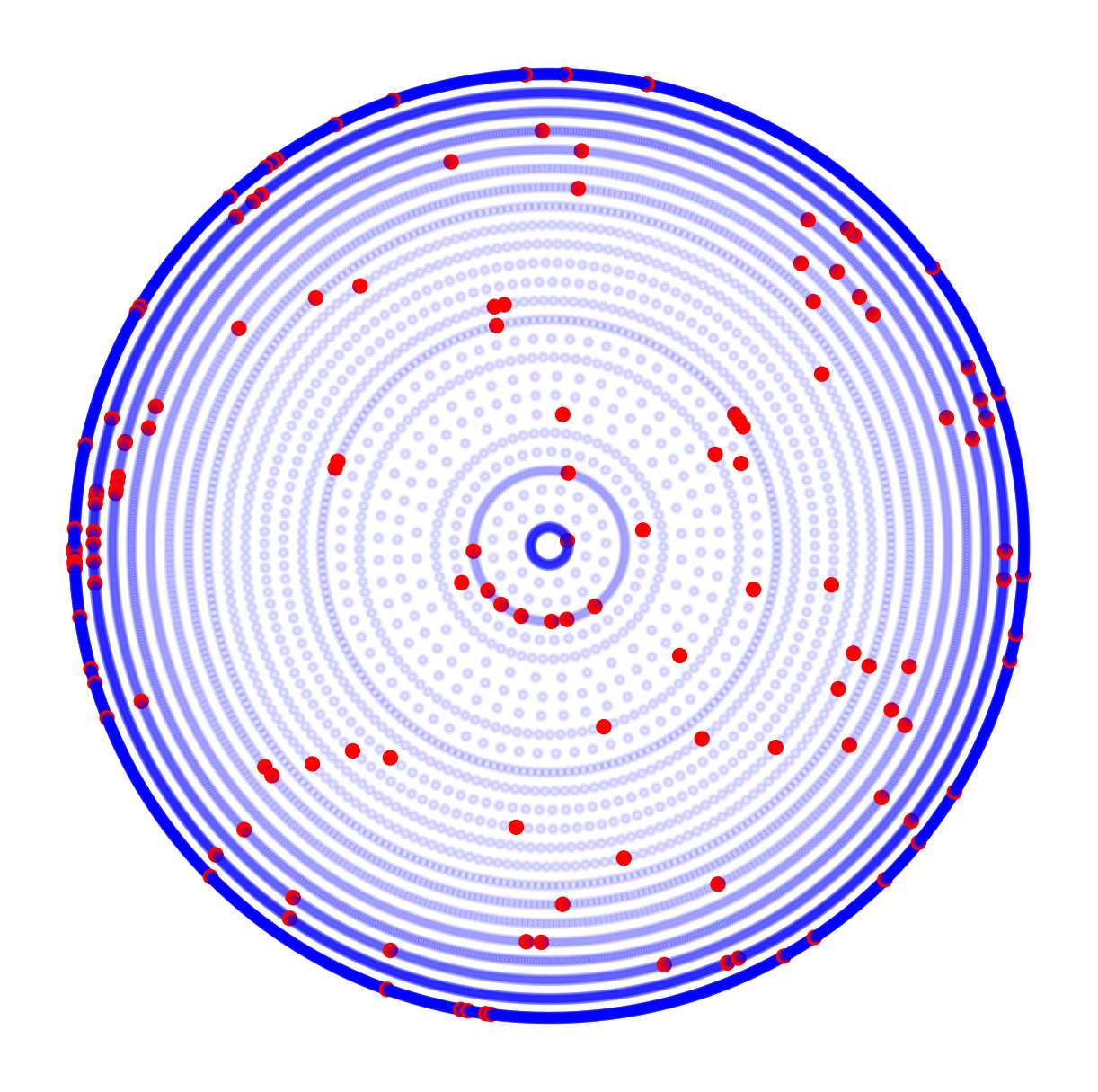} &
         \includegraphics[width=0.33\textwidth]{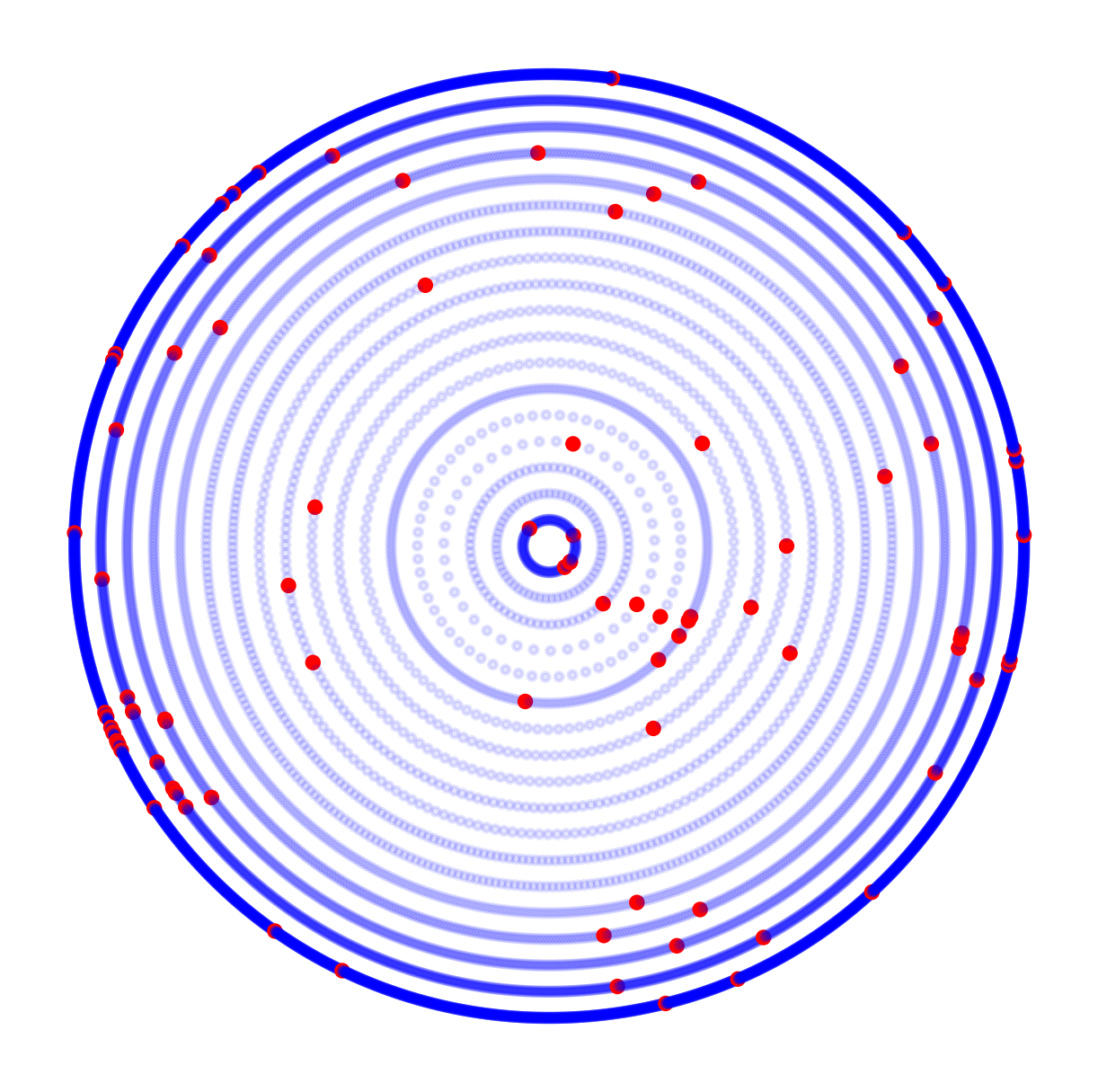}
    \end{tabular}
    \caption{K-Core Decomposition of the pre-campaign (left), electoral campaign (center), and post-election (right) networks. Larger and red-colored, resp. smaller and blue-colored, nodes indicate politicians, resp. non-politicians.}
    \label{fig:core-decomposition}
\end{figure*}

Fig.~\ref{fig:core-decomposition} illustrates the core decomposition of the networks corresponding to the three election stages. This analysis highlights the local significance of nodes across these periods and their positioning within the core or periphery of the social debate surrounding the elections. The \textit{degeneracy} values for these networks are 19, 25, and 18, respectively, suggesting that the electoral campaign phase determined the highest prevalence within the inner-most core of the network. Notably, the inner-most core of the network consistently contained at least one politician during the three observation periods, with an average core number $k$ for politicians of 5.58, 5.80, and 4.85, respectively. This result suggests that while some politicians play a pivotal ``core" role in shaping the political dialogue (especially during the campaign), this is not universally true, highlighting significant variations in local relevance. In particular, Giuseppe Conte (M5s) was consistently found in the inner-most core of all three networks, being the sole politician in this position during the first two stages. In the post-election network, he was joined in the inner-most core by Luigi Marattin (Az-Iv), Raffaella Paita (Az-Iv), Carlo Calenda (Az-Iv), and Matteo Renzi (Az-Iv).  

\begin{figure*}[t!]
    \centering
    \setlength{\tabcolsep}{0pt}
    \begin{tabular}{ccc}
         \includegraphics[width=0.33\textwidth]{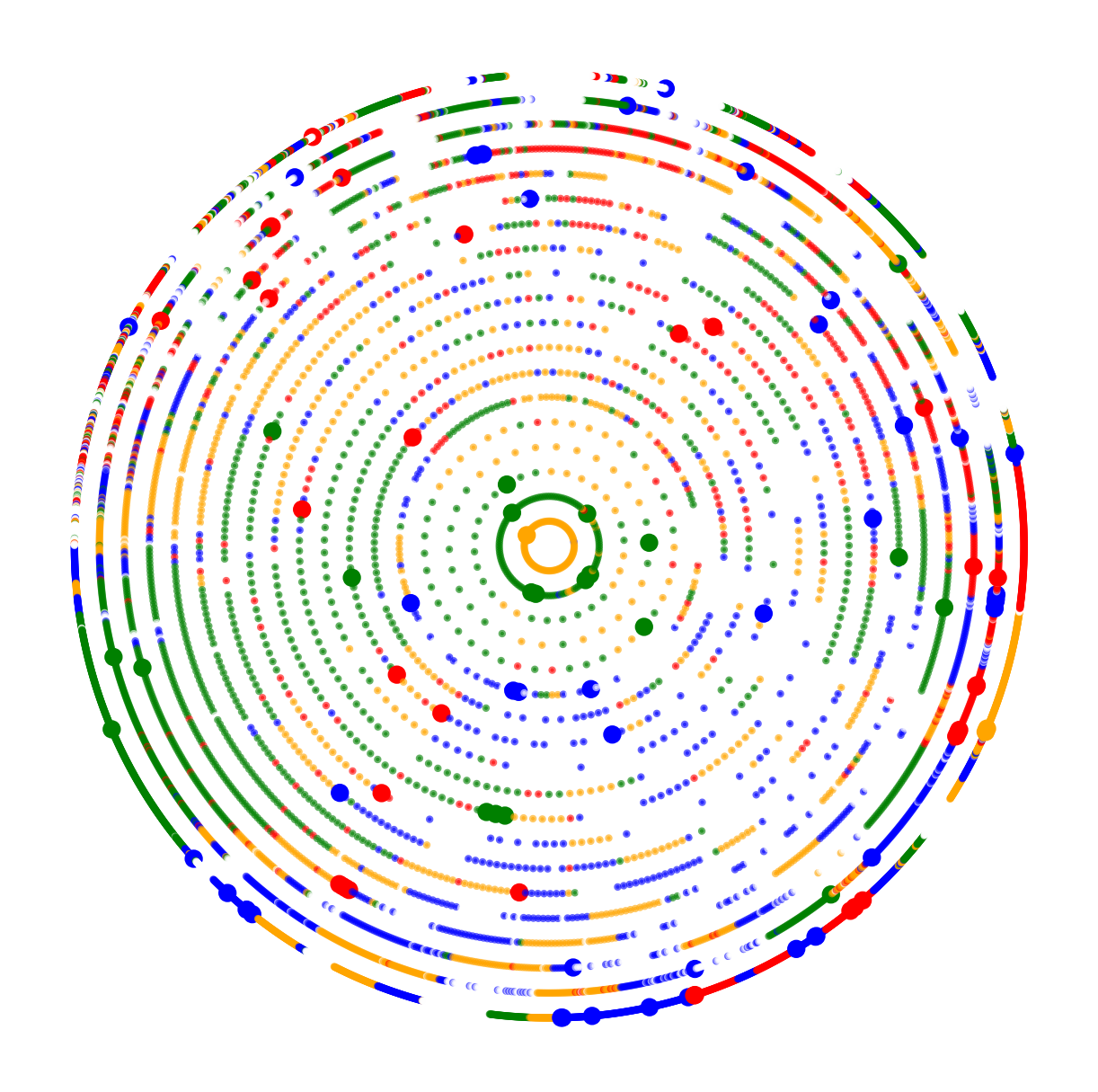} &
         \includegraphics[width=0.33\textwidth]{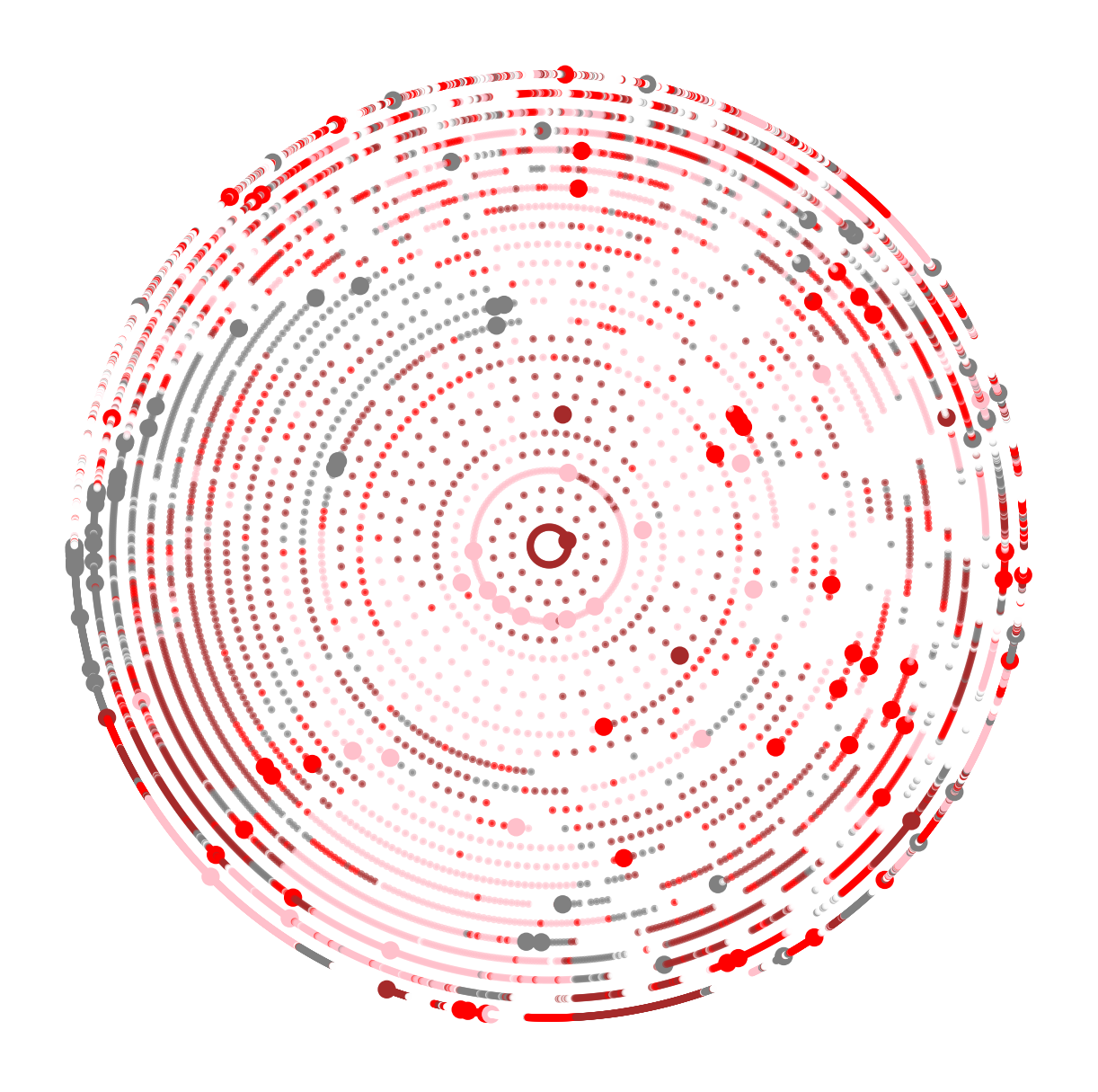} &
         \includegraphics[width=0.33\textwidth]{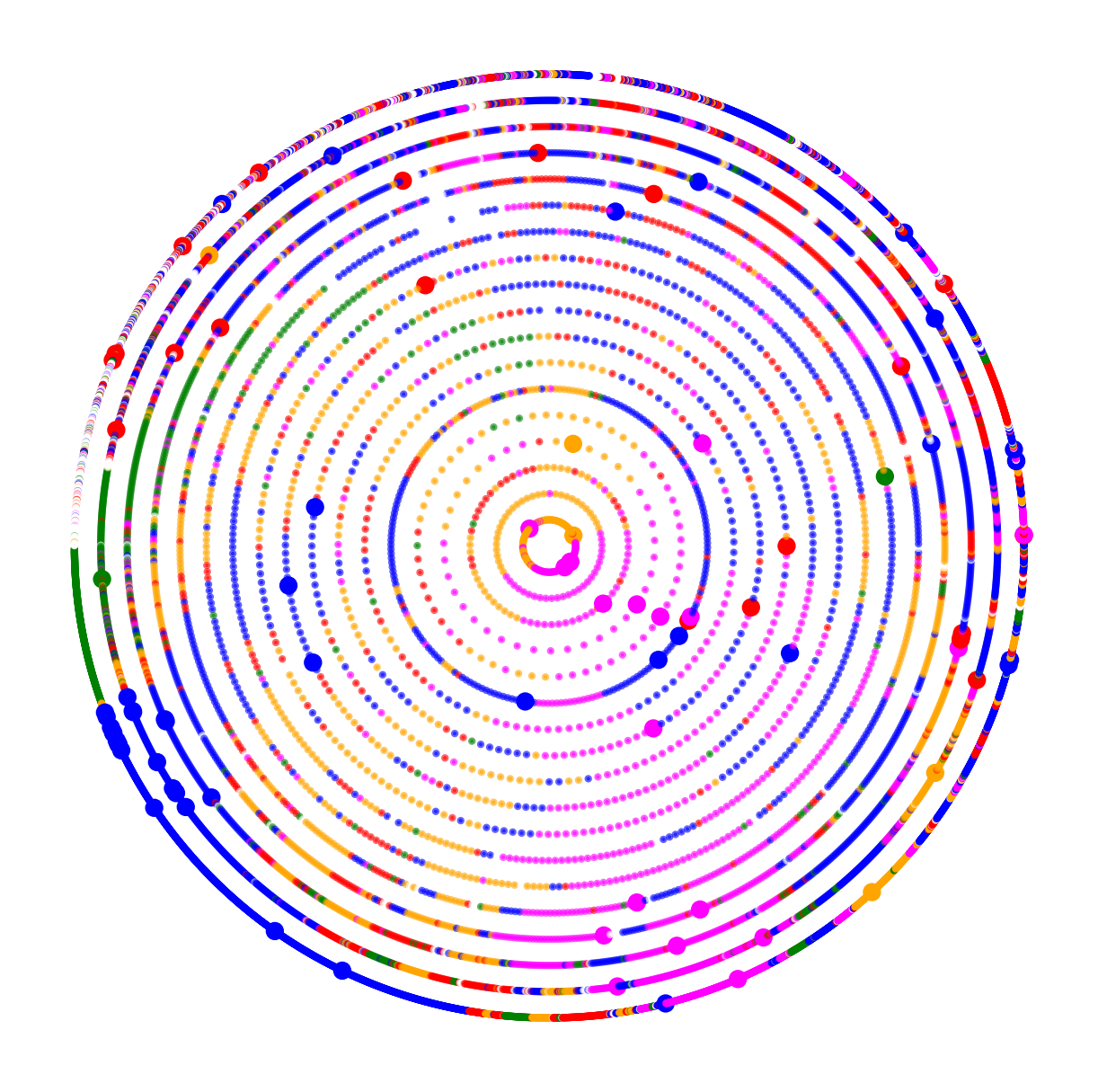}
    \end{tabular}
    \caption{K-Core Decomposition of the pre-campaign (left), electoral campaign (center), and post-election (right) networks. Colors denote political parties or coalitions, whereas larger nodes indicate politicians. For the sake of brevity, color mappings are discussed in the main text.}
    \label{fig:core-decomposition-colors}
\end{figure*}

\smallskip

We enhanced our understanding of the core decomposition of the previously mentioned networks by expanding the plots in Fig.~\ref{fig:core-decomposition} to include community membership and corresponding labels for each user or politician. This approach facilitated a more in-depth exploration of the core-periphery distribution of users over the different periods, enabling us to determine the presence of politicians within the innermost cores of the networks under consideration. Our findings are summarized in Fig.~\ref{fig:core-decomposition-colors}.

\begin{itemize}
    \item In the pre-campaign period (Fig.~\ref{fig:core-decomposition-colors}-left), politicians affiliated with the PD-AVS community (in red) dominate the network, followed by those from Lega-FdI (in blue), and Lega-Az-Iv (in green). However, in the inner-most core, only Giuseppe Conte (M5s, in orange) is represented among politicians, while the second inner-most core includes politicians primarily from the Lega-Az-Iv community, notably from Az-Iv.

    \item During the campaign period, there was a significant shift in network dynamics. The Az-Iv coalition (in pink) emerged as the most prominent, followed by the FdI-M5s coalition (in brown) and PD-AVS (in red). Interestingly, the remaining nodes form a diverse group encompassing nearly all parties (in grey), with a smaller, PD-exclusive group (also in red). As previously mentioned, despite these changes, Giuseppe Conte (M5s) continues to occupy the inner-most core, now within a broader community that also includes representatives from FdI.

    \item In the post-election network, right-wing politicians (in blue) across various coalitions (e.g., FdI-L, FdI-FI) dominate the network following the election outcome. Meanwhile, left-wing politicians from PD maintain their local significance, forming alliances with Az-Iv (in magenta) or AVS (in red). Interestingly, the election results do not significantly alter the core-periphery distribution, as politicians from M5s (in orange) continue to occupy core positions in the network. Specifically, as discussed earlier, Giuseppe Conte remains in the inner-most core along with key representatives from Az-Iv within the broader PD/Az-Iv coalition.
\end{itemize}

\subsubsection{Propaganda Usage by Political Representatives}
\label{subsec:propaganda-usage}
By applying the methodology detailed in Section~\ref{subsec:swing_voters_method}, we identified and labeled each tweet of political representatives with propaganda techniques based on their emergence at the text-fragment level. As a descriptive example, let us consider the following tweet:

\begin{quote}
    ``Il silenzio delle femministe sull'affermazione dell'UE che il velo è parte della nostra cultura è rivoltante. La sinistra ci ha abituato al doppiopesismo su tutto ma questo fa ribrezzo. Arriveranno le elezioni e gli italiani vi giudicheranno."

    \smallskip

    \noindent (``The silence of feminists on the EU's assertion that the veil is part of our culture is revolting. The left has accustomed us to double standards on everything, but this is disgusting. The elections will come, and the Italians will judge you.")
\end{quote}
This tweet is split into text fragments and annotated as follows:

\begin{quote}
    ``Il silenzio delle femministe sull'affermazione dell'UE che il velo é parte della nostra cultura é rivoltante."("The silence of feminists on the EU's assertion that the veil is part of our culture is revolting.")
    \begin{itemize}
        \item [$\cdot$] \textit{Appeal to hypocrisy, doubt.}
    \end{itemize}

   \noindent  ``La sinistra ci ha abituato al doppiopesismo su tutto ma questo fa ribrezzo." ("The left has accustomed us to double standards on everything, but this is disgusting.")

   \begin{itemize}
        \item [$\cdot$] \textit{Conversation killer, questioning the reputation.}
    \end{itemize}

   \noindent ``Arriveranno le elezioni e gli italiani vi giudicheranno." (``The elections will come, and the Italians will judge you.")

   \begin{itemize}
        \item [$\cdot$] \textit{Appeal to popularity, conversation killer, slogans.}
    \end{itemize}
\end{quote}

\smallskip

\begin{figure}[t!]
    \centering
    \includegraphics[width=.7\columnwidth]{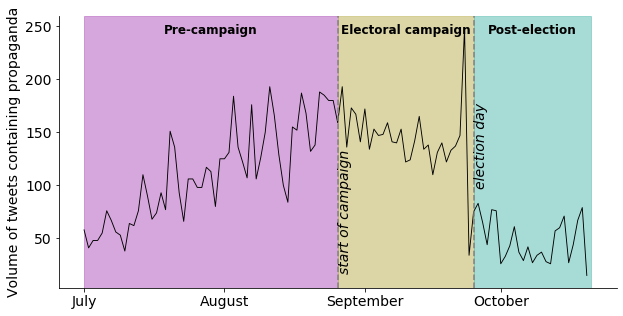}
    \caption{Number of propaganda tweets by political representatives over time (2022).}
    \label{fig:volume_propaganda}
\end{figure}

\begin{figure}[b!]
    \centering
    \includegraphics[width=.7\columnwidth]{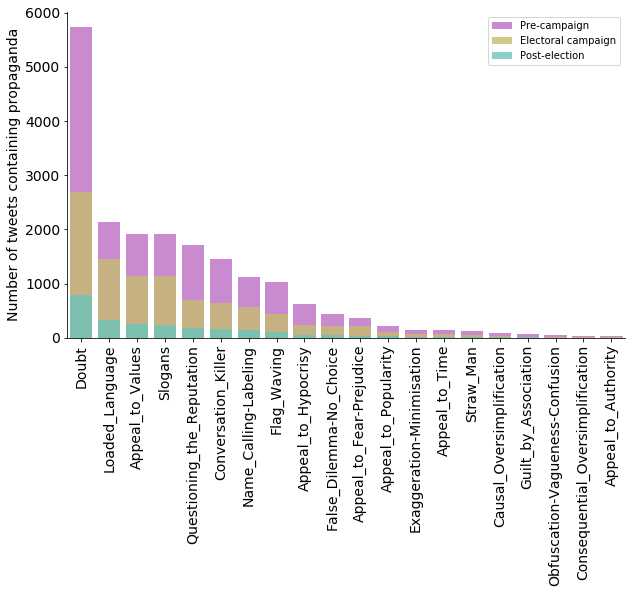}
    \caption{Cumulative number of propaganda tweets by political representatives across the three time periods, categorized by propaganda technique.}
    \label{fig:volume_tech}
\end{figure}

\noindent Given the set of annotated political representatives' tweets, we analyzed the distribution of political propaganda use over time. Fig.~\ref{fig:volume_propaganda} illustrates this distribution, revealing an increasing trend in propaganda posts before the pre-election campaign, a peak right before the election day, and a subsequent decline immediately afterward.
We then focused on the specific propaganda techniques employed by political representatives over time (cf. Fig.~\ref{fig:volume_tech}). Our analysis revealed that \textit{doubt} is the most commonly used propaganda technique, followed by \textit{loaded language}, \textit{appeal to values}, and \textit{slogans}. Notably, the variation in the use of \textit{loaded language} from the pre-campaign period to the campaign period is less pronounced than the variation in the use of \textit{doubt}.

\smallskip

\begin{figure}[b!]
    \centering
    \includegraphics[width=.95\columnwidth]{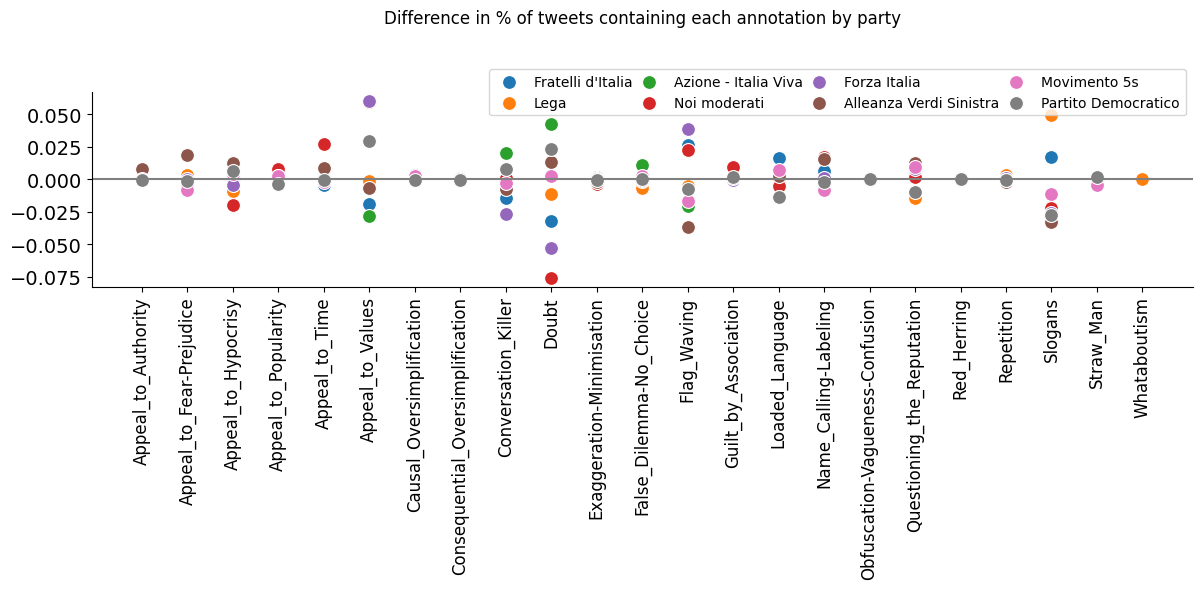}
    \caption{Use of propaganda techniques vs. baseline by political party.}
    \label{fig:party_tech}
\end{figure}

Considering the presence of representatives from several political parties, we also examined their different use of propaganda techniques. As a baseline, we considered the average frequency of use across the entire dataset of political representatives. As shown in Fig.~\ref{fig:party_tech}, some techniques exhibit distinct usage patterns across parties. Focusing on the four techniques with the most significant deviations from the baseline:

\begin{itemize}
    \item \textit{Doubt} appears to be less frequently used than the baseline by representatives from NM, FI, FdI, and Lega, all of whom are part of the right-wing coalition. In contrast, it is more commonly employed in messages from Az-Iv, PD, and AVS, which are politically placed in the center-left of the political compass. 

    \item \textit{Slogans} are used more frequently by Lega and FdI representatives compared to the baseline.

    \item \textit{Appeal to values} is more often found in the tweets of representatives from FI and PD.

    \item \textit{Flag waving} appears more common in the tweets from FI, FdI, and NM, while it is less frequent in the tweets from AVS, Az-Iv, and M5s.
\end{itemize}

\paragraph{Focus on Relevant Politicians} Given the emergence of some popular political representatives throughout the election period, we assessed the evolution in the use of propaganda by these political figures. Specifically, we focused on the representatives who exhibited significant variation in centrality (measured through PageRank) and/or user engagement (measured through in-strength) on the whole network across the three periods of interest. By analyzing these representatives, we aim to gain insights into how changes in their popularity levels relate to their use of different propaganda techniques, both in terms of type and frequency.

\smallskip

Focusing on the variations in the ranking of the used techniques between the pre-election campaign and the campaign periods, we observed that:

\begin{itemize}
    \item Luigi Marattin (Az-Iv) and Davide Faraone (Az-Iv) exhibited increased levels of both centrality and engagement from the pre-campaign to the campaign period. Interestingly, their tone of voice changed significantly between these two periods. In more detail:
    
    \begin{figure}[b!]
        \centering

        \begin{subfigure}{.45\columnwidth}
            \includegraphics[width=\linewidth]{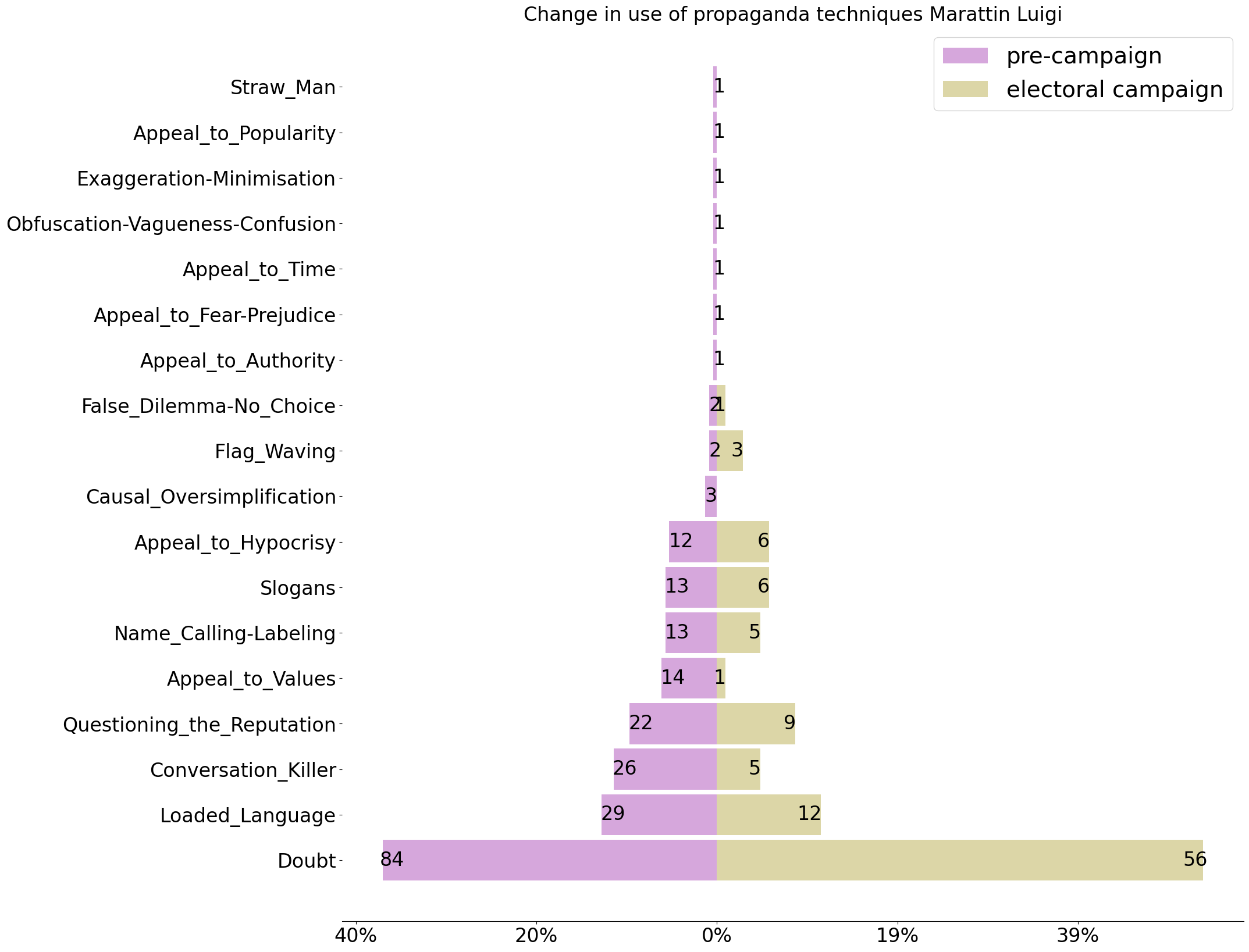}
            \caption{Luigi Marattin (Az-Iv).}
            \label{fig:marattin_bd}
        \end{subfigure}
        \hspace{1cm}
        \begin{subfigure}{.45\columnwidth}
            \includegraphics[width=\linewidth]{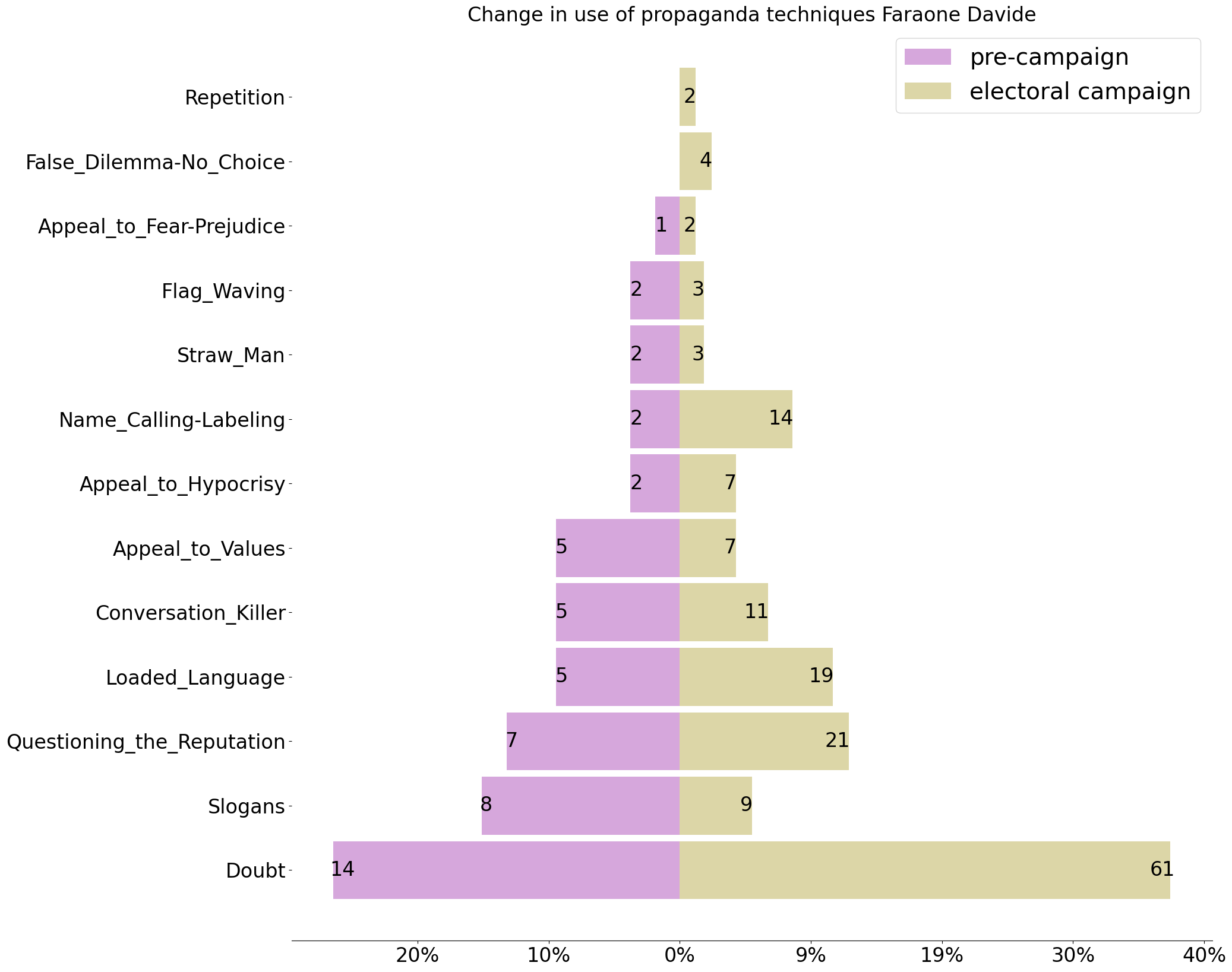}
            \caption{Davide Faraone (Az-Iv).}
            \label{fig:faraone_bd}
        \end{subfigure}

        \begin{subfigure}{.45\columnwidth}
            \includegraphics[width=\linewidth]{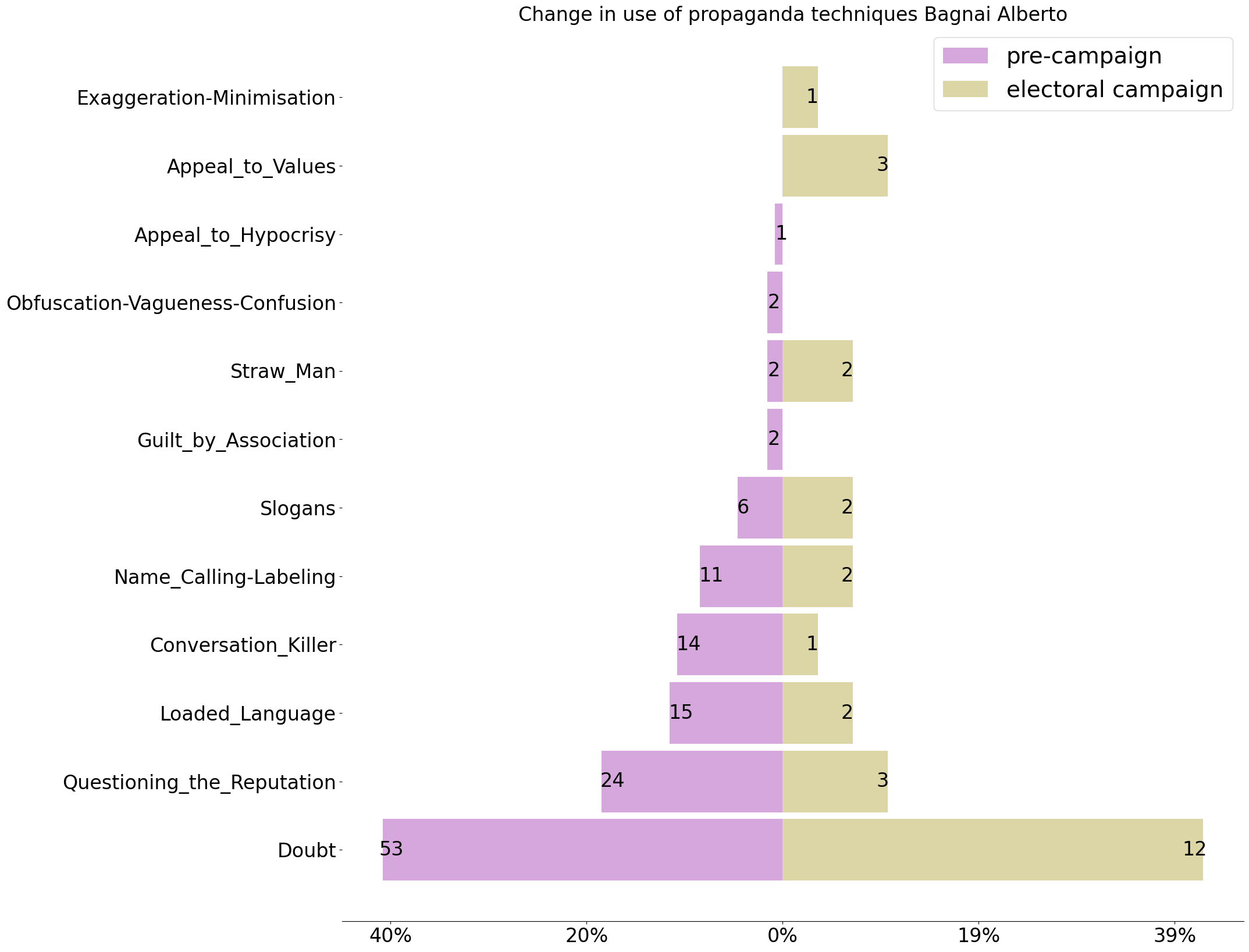}
            \caption{Alberto Bagnai (Lega).}
            \label{fig:bagnai_bd}
        \end{subfigure}
        \hspace{1cm}
        \begin{subfigure}{.45\columnwidth}
            \includegraphics[width=\linewidth]{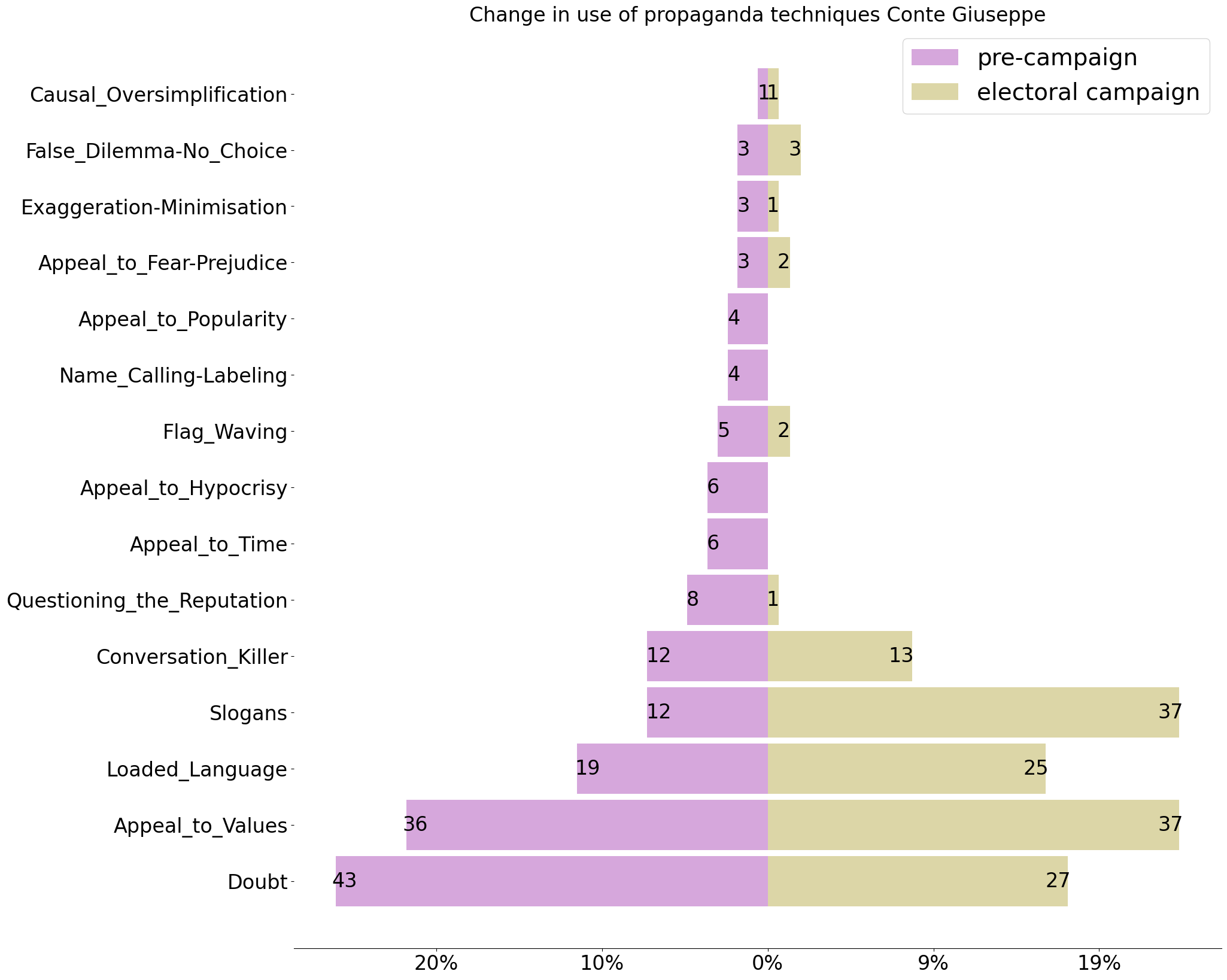}
            \caption{Giuseppe Conte (M5s).}
            \label{fig:conte_bd}
        \end{subfigure}

        \caption{Percentage change in the use of propaganda techniques pre-campaign and during the campaign by different politicians. The x-axis represents the percentage of the total vulnerability volume assigned to each propaganda technique used by the political representatives during each time period. The numbers on the bars indicate the absolute counts for each propaganda technique and time period. This framing will also be applied to the following plots.}
    \end{figure}
    \begin{itemize}
        \item As shown in Figure~\ref{fig:marattin_bd}, Marattin increased the use of \textit{doubt}, while decreasing the use of \textit{conversation killer} and \textit{appeal to values} propaganda techniques.

        \item As depicted in Figure \ref{fig:faraone_bd}, Davide Faraone also showed an increased use of \textit{doubt} and a decrease in \textit{appeal to values}. Additionally, Faraone exhibited a higher level of \textit{name calling/labeling} and a lower level of \textit{slogans}.

    \end{itemize}
    These changes suggest a strategic adjustment in their propaganda techniques, possibly aiming to enhance their influence and engagement during the campaign period by leveraging more aggressive and doubt-inducing rhetoric.

    \smallskip

    \item Alberto Bagnai (Lega) and Giuseppe Conte (M5S) experienced a notable reduced centrality and engagement between the pre-election campaign and the campaign periods\footnote{Conte only experienced a reduction in terms of centrality values.}. Specifically,

        \begin{itemize}
            \item As shown in Figure~\ref{fig:bagnai_bd}, Alberto Bagnai exhibited a lower usage level of \textit{questioning the reputation}, \textit{loaded language}, and \textit{conversation killer} during the election campaign compared to the pre-election period.

            \item As illustrated in Figure~\ref{fig:conte_bd}, Giuseppe Conte increased the use of \textit{slogans} and \textit{loaded language} while reducing the use of \textit{doubt} during the campaign period.
        \end{itemize}
    These shifts indicate that both politicians adjusted their propaganda techniques, with Bagnai adopting a less confrontational approach and Conte focusing on a more persuasive and emotionally charged rhetoric.
\end{itemize}

\noindent Focusing on the variations between the election campaign and the post-election periods, we noted that:

\begin{itemize}
    \item Matteo Renzi (Az-Iv) stood out as characterized by increasing centrality and decreasing engagement, while Enrico Letta (PD) became more central and engaging over time. In more detail,
        
        \begin{itemize}
            \item As shown in Figure~\ref{fig:renzi_da}, Renzi increased his use of \textit{conversation killer} while reducing his use of \textit{doubt}, \textit{slogans}, and \textit{appeal to hypocrisy} propaganda techniques. This shift suggests a strategic pivot to more confrontational and decisive messaging. 

            \item As illustrated in Figure~\ref{fig:letta_da}, Letta was more prone to express \textit{doubt} and use \textit{slogans}, while reducing his use of \textit{appeal to values} and \textit{loaded language}. This drift indicates a focus on fostering uncertainty and using catchy, persuasive language. 
        \end{itemize}

    \noindent Overall, these adjustments in propaganda techniques highlight the dynamic strategies employed by Renzi and Letta to navigate the changing political landscape, with each adapting their messaging to boost their online presence. 

    \smallskip

    \begin{figure}[t!]
        \centering

        \begin{subfigure}{.45\columnwidth}
            \includegraphics[width=\linewidth]{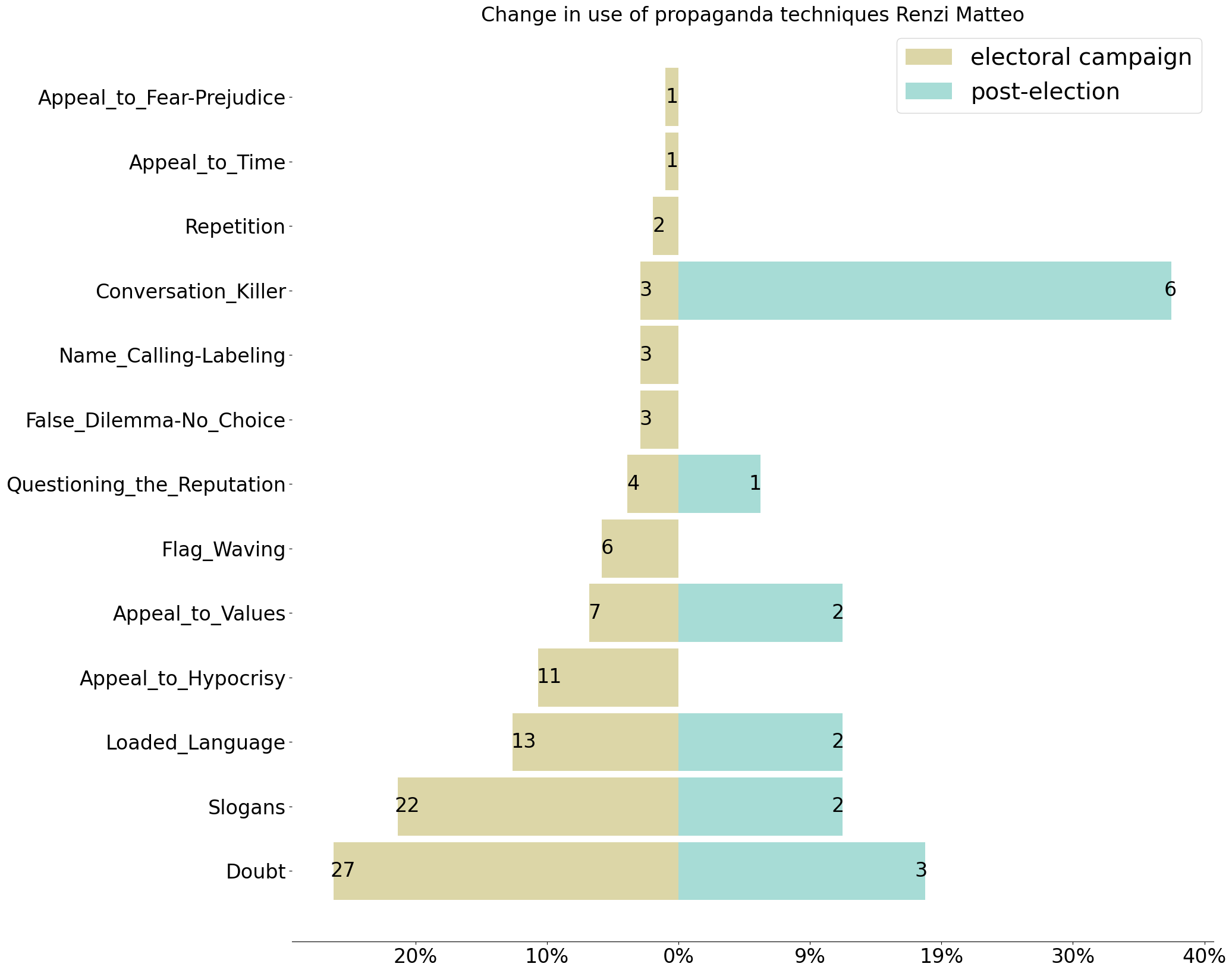}
            \caption{Matteo Renzi (Az-Iv).}
            \label{fig:renzi_da}
        \end{subfigure}
        \hspace{1cm}
        \begin{subfigure}{.45\columnwidth}
            \includegraphics[width=\linewidth]{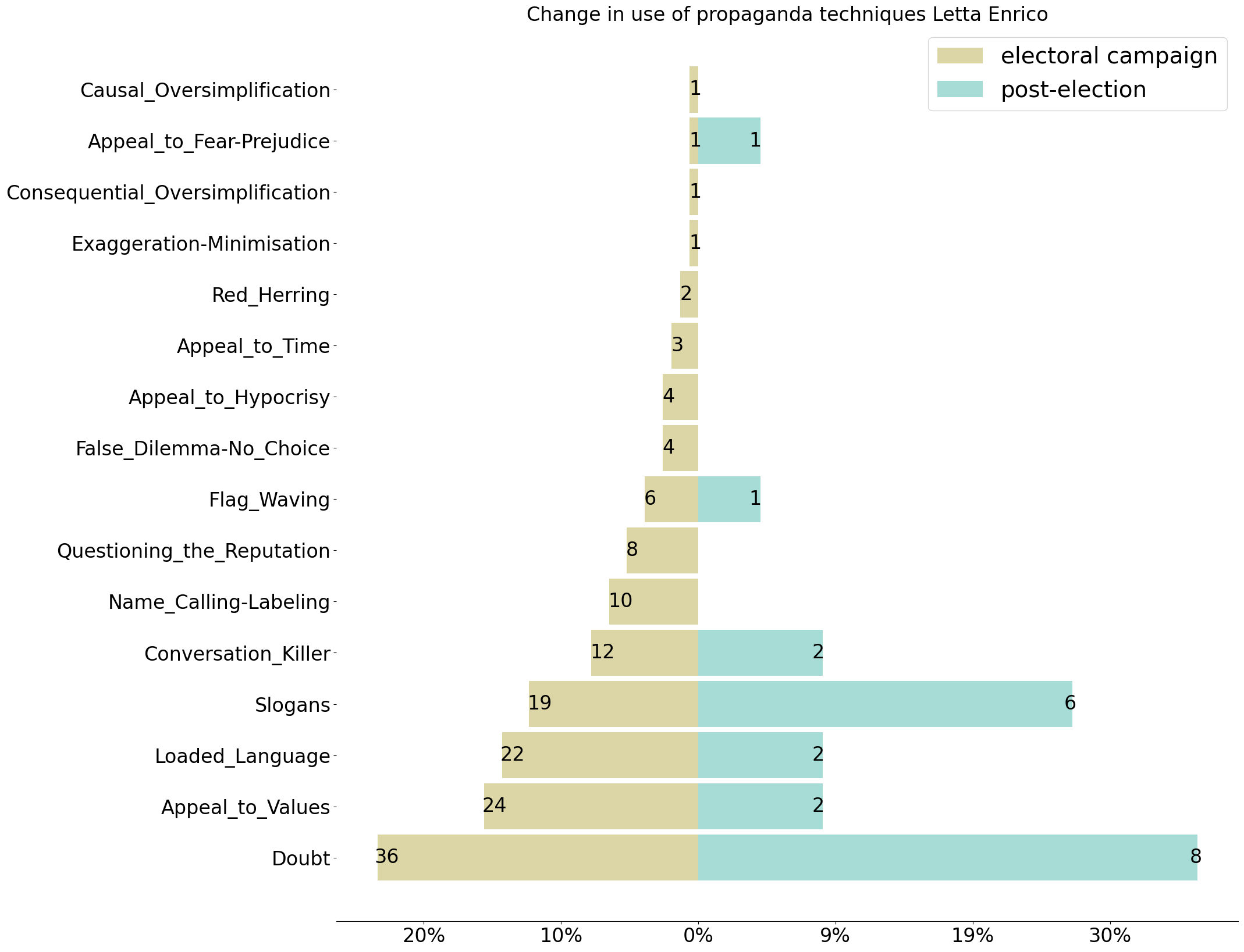}
            \caption{Enrico Letta (PD).}
            \label{fig:letta_da}
        \end{subfigure}

        \begin{subfigure}{.45\columnwidth}
            \includegraphics[width=\linewidth]{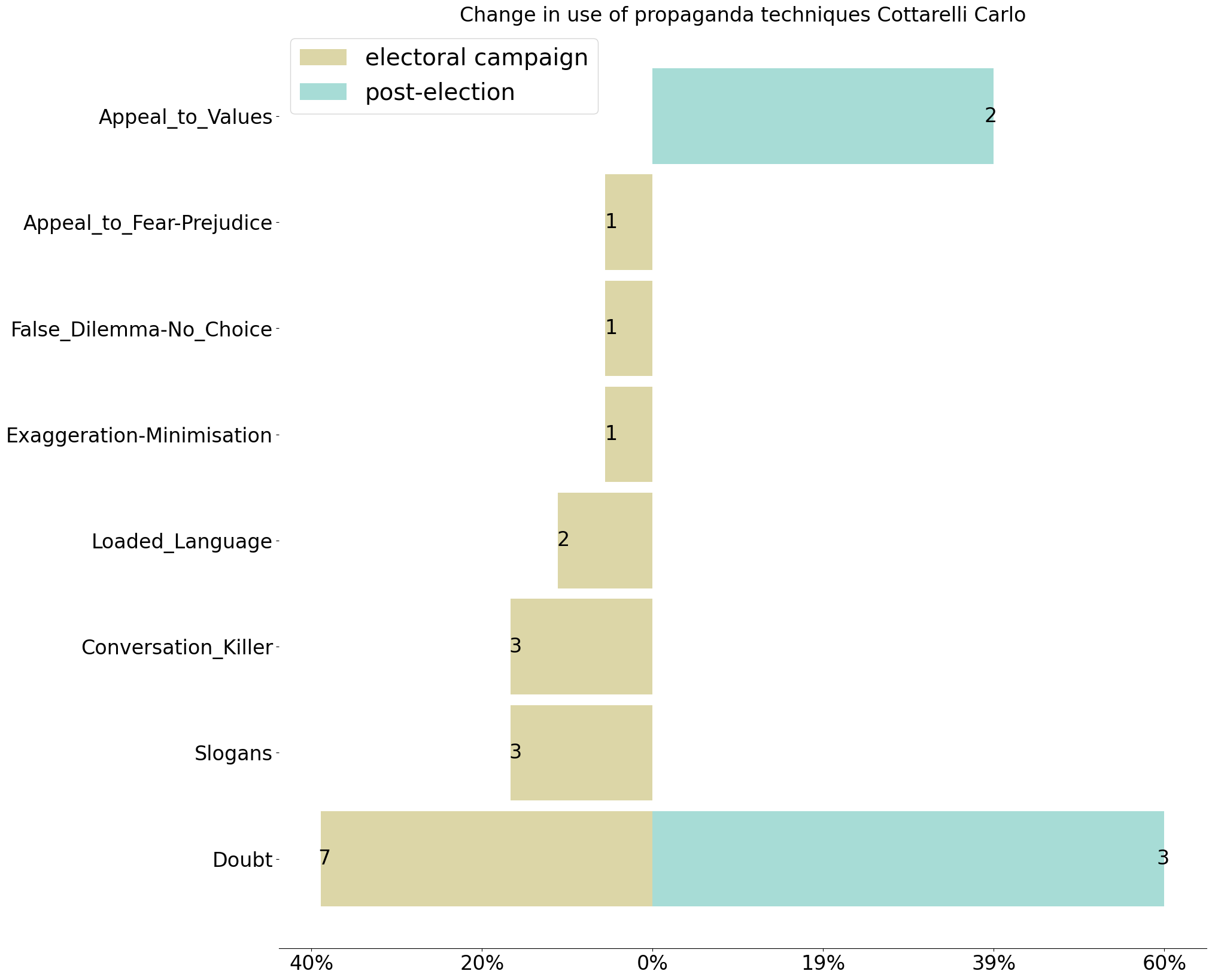}
            \caption{Carlo Cottarelli (PD).}
            \label{fig:cottarelli_da}
        \end{subfigure}
        \hspace{1cm}
        \begin{subfigure}{.45\columnwidth}
            \includegraphics[width=\linewidth]{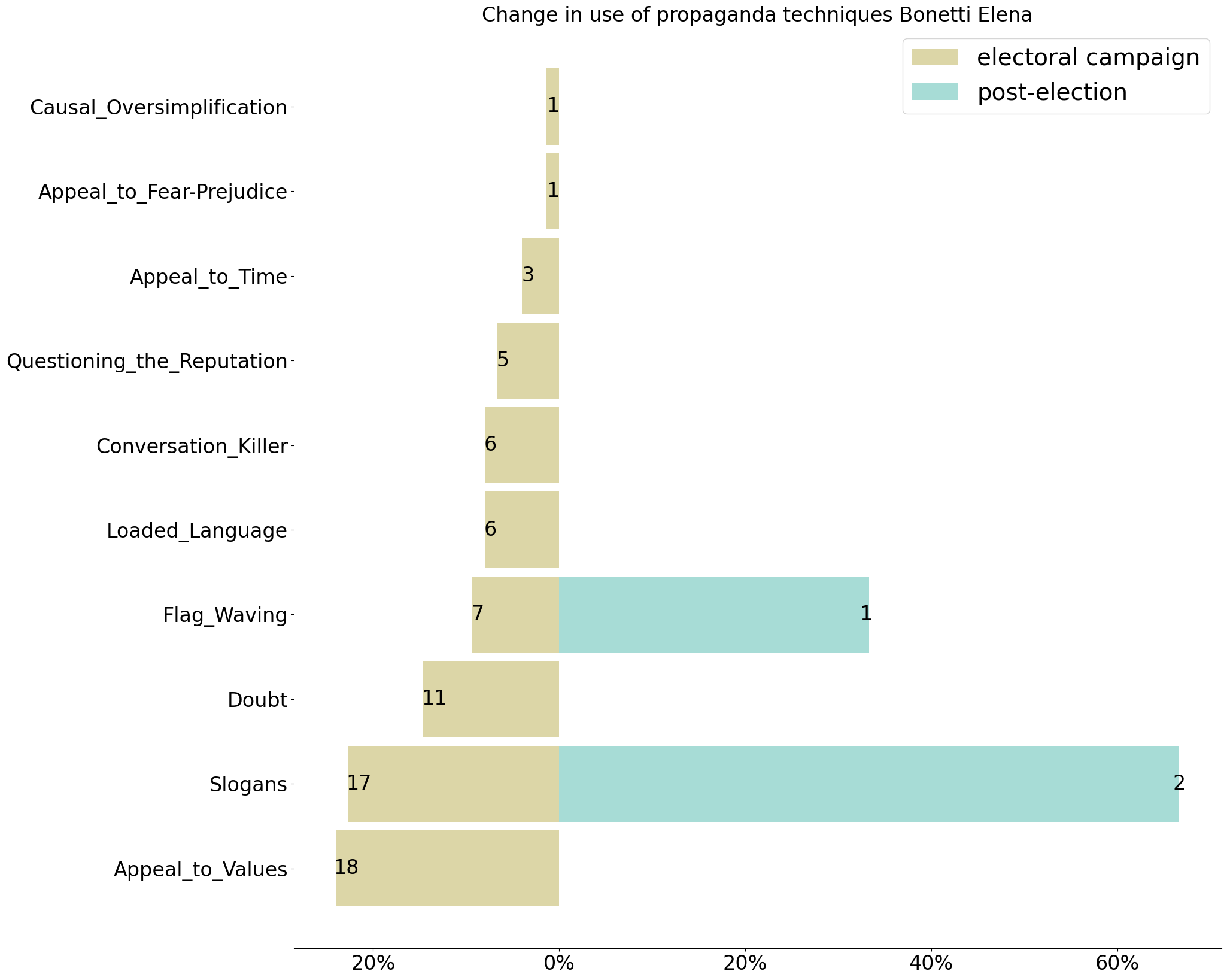}
            \caption{Elena Bonetti (Az-Iv).}
            \label{fig:bonetti_da}
        \end{subfigure}

        \caption{Percentage change in the use of propaganda techniques during-campaign and post-election by different politicians.}
    \end{figure}

    \item Carlo Cottarelli (PD) and Elena Bonetti (Az-Iv) experienced a decrease in popularity post-election, specifically in terms of centrality for Cottarelli and in both centrality and engagement for Bonetti. In terms of propaganda techniques:

    \begin{itemize}
        \item As shown in Figure~\ref{fig:cottarelli_da}, Carlo Cottarelli increased his use of \textit{doubt} and \textit{appeal to values} over time, while decreasing the use of \textit{slogans} and \textit{conversation killer}. This change indicates a shift towards more skeptical and value-oriented messaging.

        \item As visible in Figure~\ref{fig:bonetti_da}, Elena Bonetti increased her use of \textit{slogans} and \textit{flag waving} as well as the volume of \textit{appeal to values} and \textit{doubt}. This drift suggests an attempt to boost engagement through emotionally resonant and patriotic messaging. 
    \end{itemize}

    \noindent Once again, these shifts in propaganda techniques reflect the efforts of politicians to adapt to their changing political influence, with each representative modifying their communication strategies to address their evolving roles and audience dynamics in the post-election period.
\end{itemize}

\noindent Finally, we assessed variations in the use of propaganda techniques between the pre-campaign and the post-election phases.

\begin{itemize}
    \item In this case, we focused on two political representatives characterized by different variations in engagement: Matteo Salvini, who had an increasing in-strength post-election, and Giorgia Meloni, who had a decreasing in-strength despite being the leader of the winning party. In particular, 

    \begin{itemize}
        \item As from Figure~\ref{fig:salvini_ba}, we can note how Matteo Salvini began to use more \textit{doubt} and \textit{name calling/labeling} and less \textit{slogans} and \textit{conversation killer} propaganda techniques post-election. This change indicates a strategic shift towards more aggressive and skeptical messaging to enhance popularity after the election.
    
        \smallskip
    
        \item As depicted in Figure \ref{fig:meloni_ba}, Giorgia Meloni tended to use more \textit{appeal to values} and \textit{flag waving} and less \textit{questioning the reputation} and \textit{slogans} post-election. This drift suggests a focus on patriotic and value-oriented messaging to maintain support and influence on the online platform.
    
        \begin{figure}[t!]
            \centering
    
            \begin{subfigure}{.45\columnwidth}
                \includegraphics[width=\linewidth]{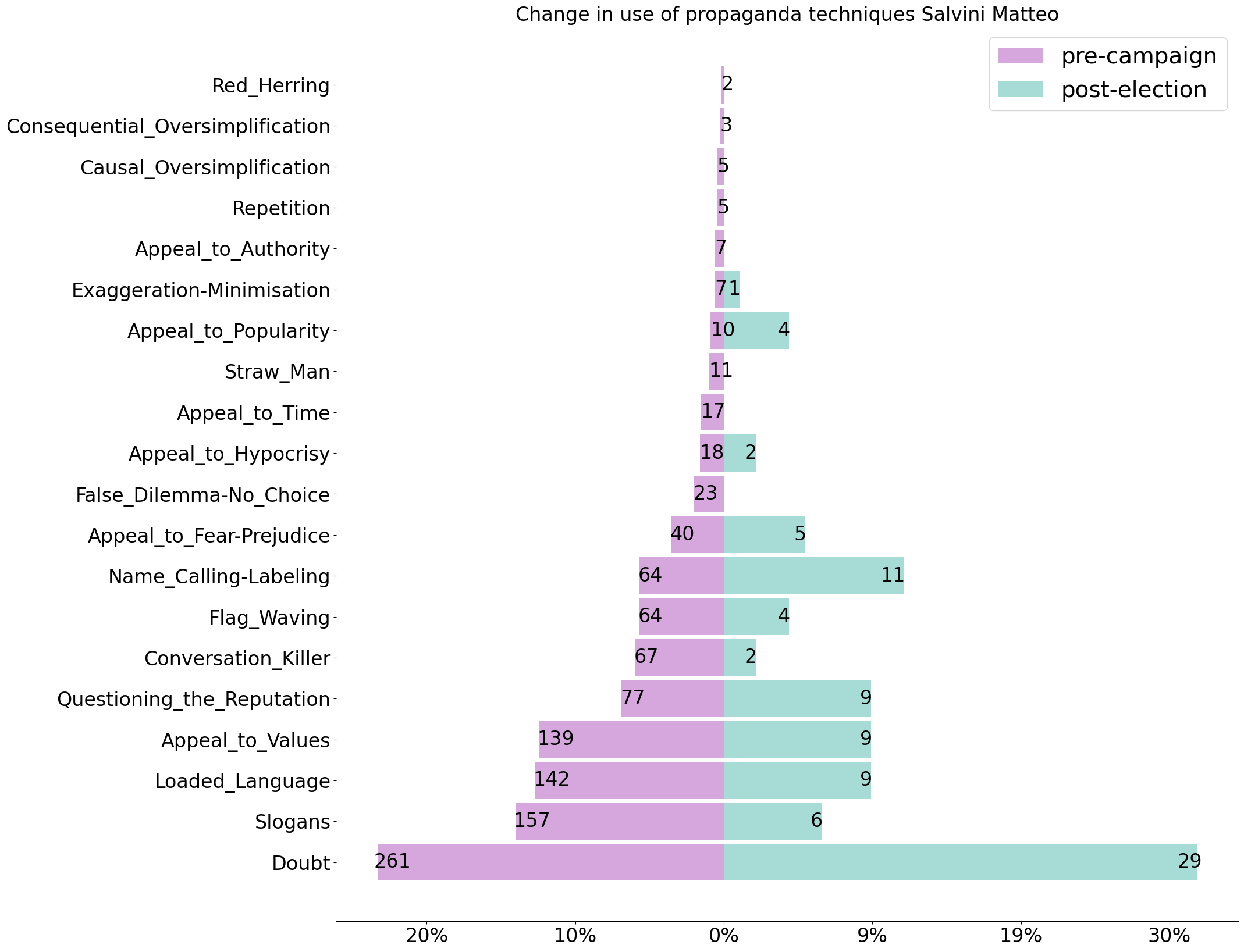}
                \caption{Matteo Salvini (Lega).}
                \label{fig:salvini_ba}
            \end{subfigure}
            \hspace{1cm}
            \begin{subfigure}{.45\columnwidth}
                \includegraphics[width=\linewidth]{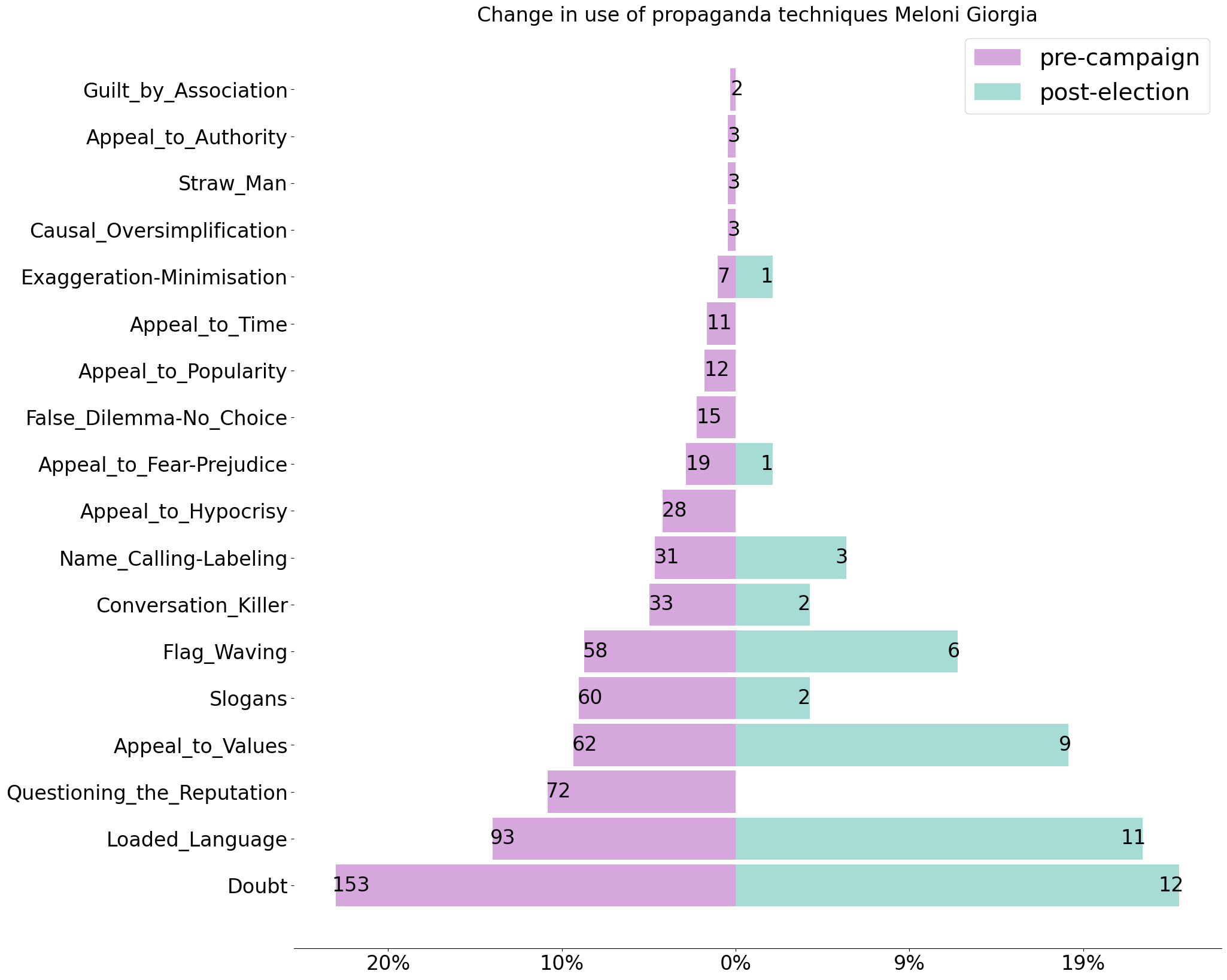}
                \caption{Giorgia Meloni (FdI).}
                \label{fig:meloni_ba}
            \end{subfigure}
    
            \caption{Percentage change in the use of propaganda techniques pre-campaign and post-elections by Matteo Salvini and Giorgia Meloni.}
            \label{fig:partytonan_both}
        \end{figure}
    \end{itemize}
    
    \noindent These variations in propaganda techniques highlight how Salvini and Meloni adjusted their posting strategies and reflect two different approaches to maintaining or increasing their political influence in the post-election period.
\end{itemize}

\subsection{Swing Voters Detection and Vulnerability to Propaganda Techniques}\label{subsec:swing_voters_results}

In the following, we discuss the swing voters identified in our dataset, detailing the most common types of political shifts they undergo. We then examine the susceptibility of these users to specific types of political propaganda and analyze whether particular categories of swing voters are more vulnerable to certain propaganda techniques.

\subsubsection{Swing Voters Detection}\label{subsubsec:swing_voters_detection_results}

In Section~\ref{subsubsec:community_analysis}, we examined the evolution of the largest communities over the three periods of interest. Our analysis revealed that the political discourse on the social platform remained relatively stable, with most users either staying within their original communities or migrating to other communities that aligned with their existing political views. However, a closer look at user transitions uncovers some notable exceptions to this general trend.

\smallskip

In particular, considering the period before and during the electoral campaign, we can note that 
\begin{itemize}
    \item 7.62\% of users were identified as hard swing voters (576 out of 7,564 active users in both periods). Interestingly, most transitions occurred from right-wing to left-wing affiliations. Specifically, many users transitioned from FdI to PD (108 users) and from [L, Az-Iv] to [AVS, PD] (100 users), while another significant group transitioned from [AVS, PD] to Az-Iv and [M5S, FdI] (76 and 64 users, respectively).

    \smallskip

    \item 5.05\% of users (382 out of 7,564 active users in both periods) moved from a political to a non-political community between the two periods. Most of these users shifted from the FdI right-wing party or the right-wing coalitions [L, FdI] and [L, Az-Iv] to non-political communities. Other users migrated from M5S, with a minority coming from the left-wing community [AVS, PD].

    \smallskip

    \item 2.78\% (210 out of 7,564 active users in both periods) moved from a non-political to a political community, with most of these users joining the right-wing coalition [FI, L, PD, FdI, NM].
\end{itemize}

\smallskip

\noindent If we focus on the period from the electoral campaign to the post-election phase, we can observe that:

\begin{itemize}
    \item Notably, 21.84\% of users (2,110 out of 9,660 active users in both periods) moved from a non-political community to a political group. The majority of these users (1,914 out of 2,110) transitioned to a right-wing coalition, which won the elections. The remaining users migrated to a left-wing coalition [AVS, PD] or to M5S.

    \smallskip

    \item 6.43\% of users (622 out of 9,660 active users in the second period) were identified as hard swing voters. A significant portion of these users (294 out of 622) transitioned from left-wing communities (PD or [AVS, PD]) to the right-wing coalition [Az-Iv, FdI]. Conversely, there were also migrations from right-wing to left-wing coalitions, with users moving from Az-Iv or [M5S, FdI] to [AVS, PD] (216 out of 622). Some users (50 out of 622) transitioned from [AVS, PD] to M5S.

    \smallskip

    \item A small minority of users (0.49\%) transitioned to a non-political community after the elections, with most of these migrations coming from [AVS, PD].
\end{itemize}

\subsubsection{Vulnerability to Propaganda}
After identifying swing voters, we focused on comparing the vulnerability of non-swing and swing voters to propaganda messages shared by political representatives. Specifically, we assessed how susceptible each group (and swing voters subgroup) was to endorsing the propaganda by examining their retweeting behavior.

\smallskip

\noindent In analyzing swing voters' vulnerability, we observed a lower tendency to retweet propaganda tweets after a political shift occurs compared to before. In particular, the ratios of propaganda retweets \textit{post over pre-shift} for each period are as follows: 0.538 electoral campaign/pre-campaign, 0.765 post election/electoral campaign, and 0.446 post election/pre-campaign.
Despite not experiencing a political endorsement change, non-swing voters also exhibit vulnerability to propaganda techniques, even though these users are less likely to retweet propaganda messages compared to swing voters. Nonetheless, the difference in the volume of propaganda retweets across subsequent periods is more pronounced among non-swing voters. Specifically, their ratios of propaganda retweets are as follows: 0.297 during campaign/pre, 0.325 after elections/during campaign, and 0.322 after elections/before campaign. Table~\ref{tab:prop_retweet} offers a summary of the volume of propaganda retweets by period.

\begin{table}[t!]
\centering

\caption{Propaganda retweets volume by period, for swingers and non-swingers. \label{tab:prop_retweet}}

{\footnotesize
    \setlength{\tabcolsep}{3pt}
    \begin{tabular}{rcccccc}
        \toprule
    
        \multicolumn{1}{c}{
            \textbf{Political shift period}
        } &  
        \multicolumn{1}{c}{
            \begin{tabular}[c]{@{}c@{}} \textbf{Swing voters}\\ \textbf{pre shift}
            \end{tabular} 
        } & 
        \multicolumn{1}{c}{
            \begin{tabular}[c]{@{}c@{}} \textbf{Swing voters}\\ \textbf{post shift}
            \end{tabular}
        } & 
        \multicolumn{1}{c}{
            \begin{tabular}[c]{@{}c@{}} \textbf{Non-swing voters}\\ \textbf{pre shift}
            \end{tabular} 
        } & 
         \begin{tabular}[c]{@{}c@{}} \textbf{Non-swing voters}\\ \textbf{post shift}
        \end{tabular} 
        \\
    
        \midrule
        
        Pre-campaign - campaign & 
        6\,300 & 
        3\,389 & 
        5\,467 &
        1\,622 
        \\

        Campaign - post election & 
        24\,290 & 
        18\,582 & 
        15\,766 &
        5\,123 
        \\

        Pre-campaign - post election & 
        41\,332 & 
        18\,419 & 
        16\,157 &
        5\,204 
        \\
        \bottomrule
    \end{tabular}
    \vspace*{-0.3truecm}
}
\end{table}



    
    
        



\smallskip

Focusing specifically on swing voters, we identified five types of users based on the characteristics of their political shift, namely hard swing voters, soft swing voters, spurious swing voters, apolitical-to-political swing voters, and political-to-apolitical swing voters (cf. Section~\ref{subsec:swing_voters_method} for definitions and details). Table~\ref{tab:swingers_type_vuln} reports the number of users vulnerable to propaganda tweets in each time period by swing type, considering pre and post political shifts.
To understand the relationship between types of swing voters and their vulnerability to propaganda techniques used by political representatives, we analyzed the volume of each propaganda technique to which a user is found to be vulnerable. Specifically, for each group of users identified as swing voters of a given type, we counted the frequency of different propaganda techniques within the tweets retweeted by these users. This approach allows us to measure how often each propaganda technique is endorsed by different types of swing voters, providing insights into their susceptibility to various forms of political messaging.

\begin{table}[b]
\centering


\caption{Number of swing voters vulnerable to propaganda tweets in each time period, considering pre and post political shift.}
\label{tab:swingers_type_vuln}

{\footnotesize
    \setlength{\tabcolsep}{5pt}
    \renewcommand{\arraystretch}{1.15}
    
    \begin{tabular}{rrrrrrr}
        \toprule


        \textbf{Swinger type}
        &  
        \multicolumn{2}{c}{
            \begin{tabular}[c]{@{}c@{}} 
                \textbf{Pre-campaign/}\\ 
                \textbf{Campaign}
            \end{tabular} 
        } &  
        \multicolumn{2}{c}{
            \begin{tabular}[c]{@{}c@{}} 
                \textbf{Campaign/}\\ 
                \textbf{Post elections}
            \end{tabular} 
        } & 
        \multicolumn{2}{c}{
            \begin{tabular}[c]{@{}c@{}} 
                \textbf{Pre-campaign/}\\ 
                \textbf{Post elections}
            \end{tabular} 
        } 
        \\

        \rowcolor{gray!10}
        \cellcolor{white}
        &  
        \multicolumn{1}{c}{\textit{Pre}} & 
        \multicolumn{1}{c}{\textit{Post}}
        &  
        \multicolumn{1}{c}{\textit{Pre}} & 
        \multicolumn{1}{c}{\textit{Post}}
        &  
        \multicolumn{1}{c}{\textit{Pre}} & 
        \multicolumn{1}{c}{\textit{Post}} 
        \\
    
        \midrule
        
        Hard swing voters & 
        77 & 
        68 & 
        172 &
        180 &
        307 &
        231 
        \\

        \rowcolor{gray!10}
        Soft swing voters & 
        1 & 
        1 & 
        0 &
        0 &
        7 &
        6 
        \\

        Spurious swing voters & 
        21 & 
        16 & 
        346 &
        391 &
        593 &
        409 
        \\

        \rowcolor{gray!10}
        Apolitical-to-political swing voters & 
        665 & 
        543 & 
        2\,128 &
        2\,245 &
        1\,667 &
        1\,492 
        \\

        Political-to-apolitical swing voters & 
        41 & 
        27 & 
        8 &
        11 &
        16 &
        16 
        \\
        \bottomrule
    \end{tabular}
    \vspace*{-0.3truecm}
}
\end{table}

\begin{itemize}
    \item Regarding \textit{hard swing voters}, notable differences in the retweet patterns of propaganda techniques emerge when comparing the pre-campaign period to the post-election period. There is a noticeable change in the ranking of the most commonly retweeted techniques. Specifically, \textit{conversation killer} and \textit{appeal to hypocrisy} decrease in rank post-elections, while \textit{loaded language}, \textit{slogans}, and \textit{name calling/labeling} gain higher popularity after election times (cf. Figure~\ref{fig:tech_ba_hard}).
    The three most common swing patterns among hard-swinging users are:
    \begin{itemize}
        \item (L, Az-Iv) $\rightarrow$ (PD, AVS);
        \item M5s $\rightarrow$ (PD, AVS);
        \item M5s $\rightarrow$ (FdI, Az-Iv).
    \end{itemize}

    \begin{figure}[t!]
        \centering
        \includegraphics[width=.7\columnwidth]{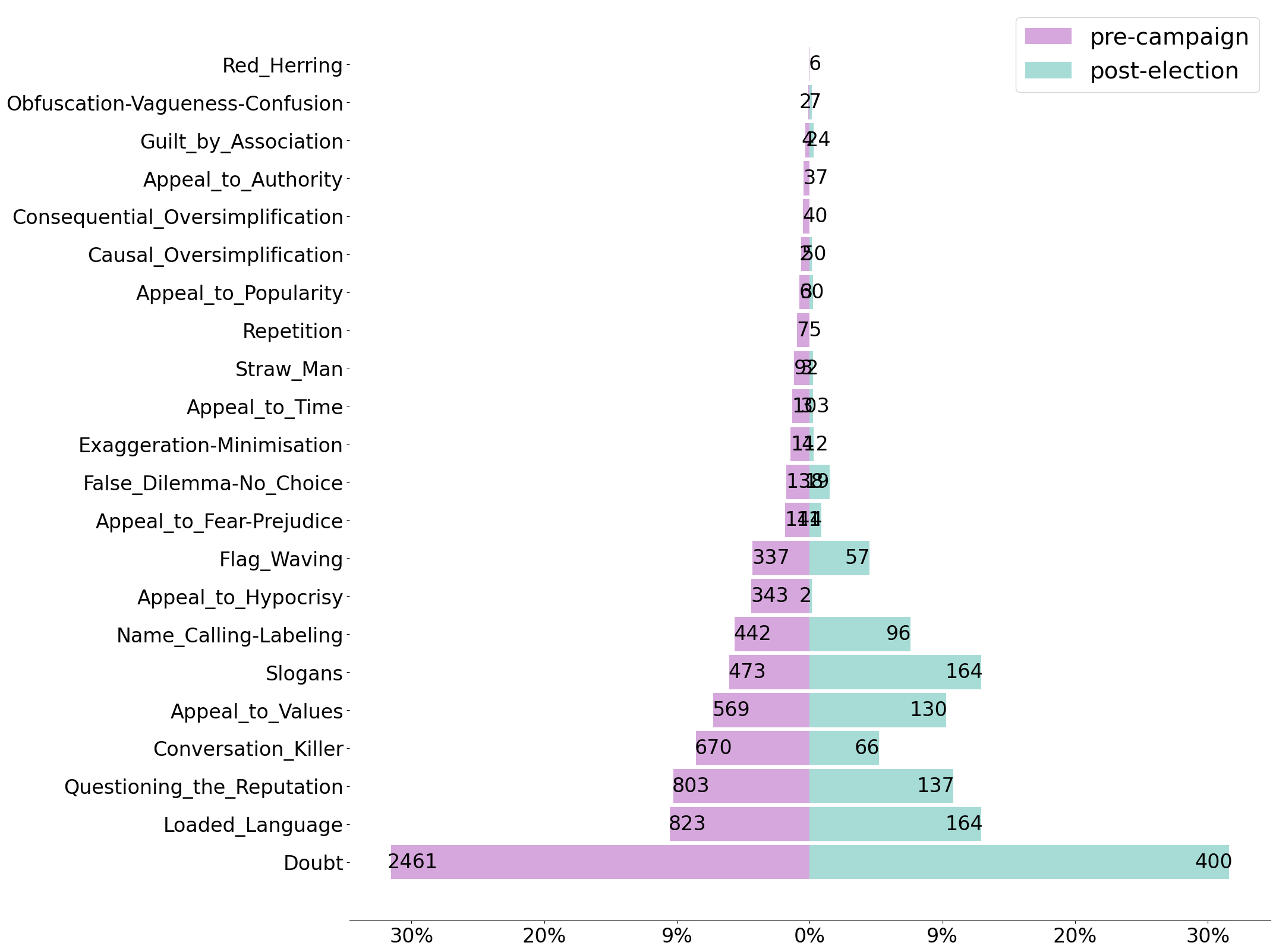}
        
        \caption{Vulnerability to propaganda techniques pre-campaign and post-election for hard swing voters. The x-axis represents the percentage of total vulnerability volume assigned to each propaganda technique for the time periods considered. The numbers on the bars indicate the absolute counts for each propaganda technique during each period. These details also apply to the following plots.}
        \label{fig:tech_ba_hard}
    \end{figure}

    \item In the group of \textit{soft swing voters}, there are also notable differences in propaganda retweeting behaviors between the pre-campaign and post-election periods. Specifically, the popularity of \textit{doubt} and \textit{name calling/labeling} tends to decrease post-elections, while \textit{conversation killer} and \textit{appeal to values} techniques gain greater popularity after election times (cf. Figure~\ref{fig:tech_ba_soft}). 
    The most common swing pattern for soft swing voters is (L, Az-Iv) $\rightarrow$ (FdI, FI).

        \begin{figure}[t!]
            \centering
            \includegraphics[width=.7\columnwidth]{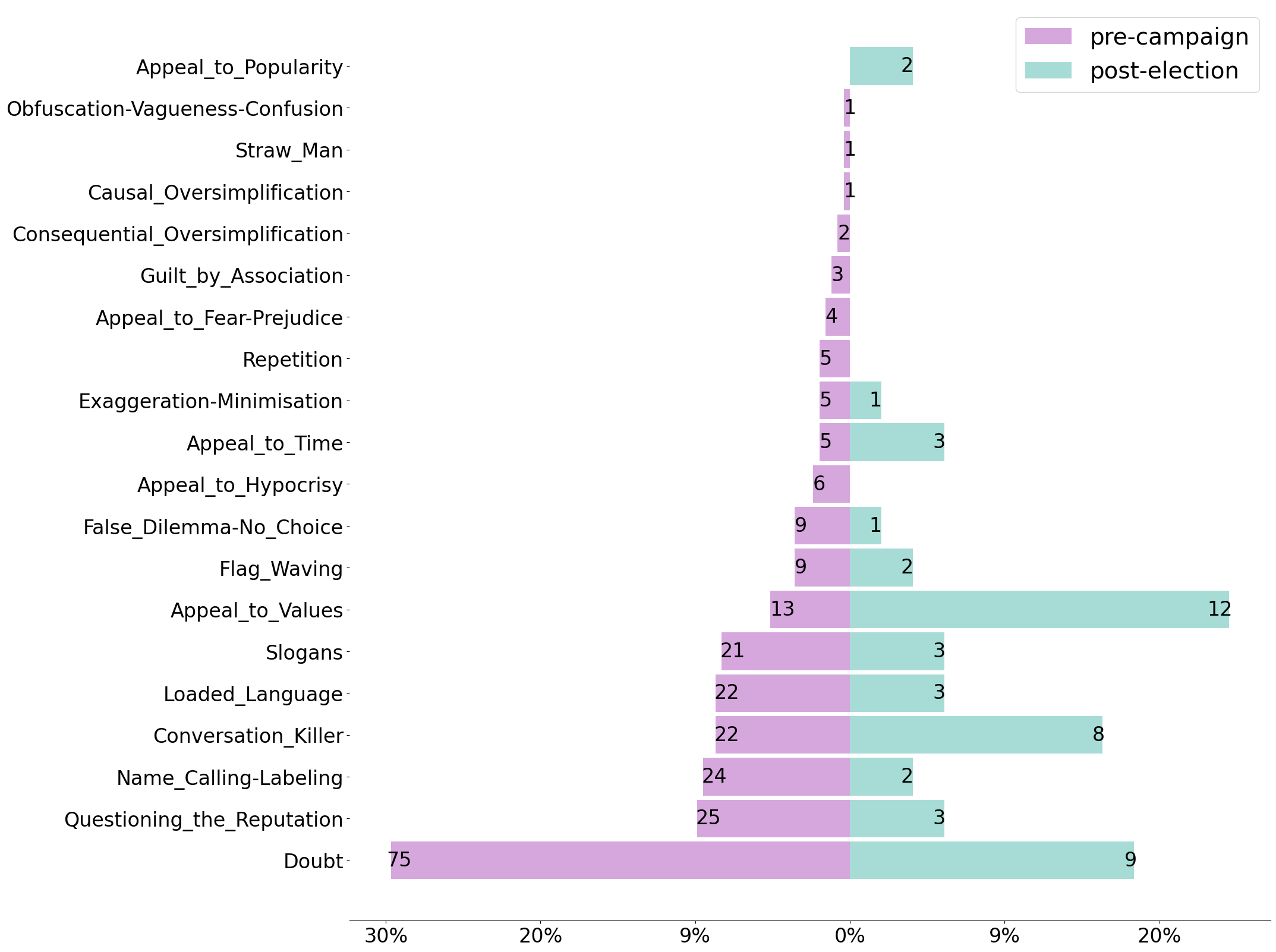}
            
            \caption{Vulnerability to propaganda techniques pre-campaign and post-election for soft swing voters.}
            \label{fig:tech_ba_soft}
        \end{figure}

    \smallskip

    \item For spurious swing voters, we observe a shift in propaganda retweeting patterns between the campaign and post-election periods. In particular, there is an increase in the use of \textit{appeal to values} and a decrease in the use of \textit{slogans} (cf. Figure \ref{fig:tech_da_noreal}). The most common swing patterns among these voters are:
    
        \begin{itemize}
            \item Az-Iv $\rightarrow$ (PD, Az-Iv);
            \item (FdI, M5s) $\rightarrow$ M5s.
        \end{itemize}
    
        \begin{figure}[t!]
            \centering
            \includegraphics[width=.7\columnwidth]{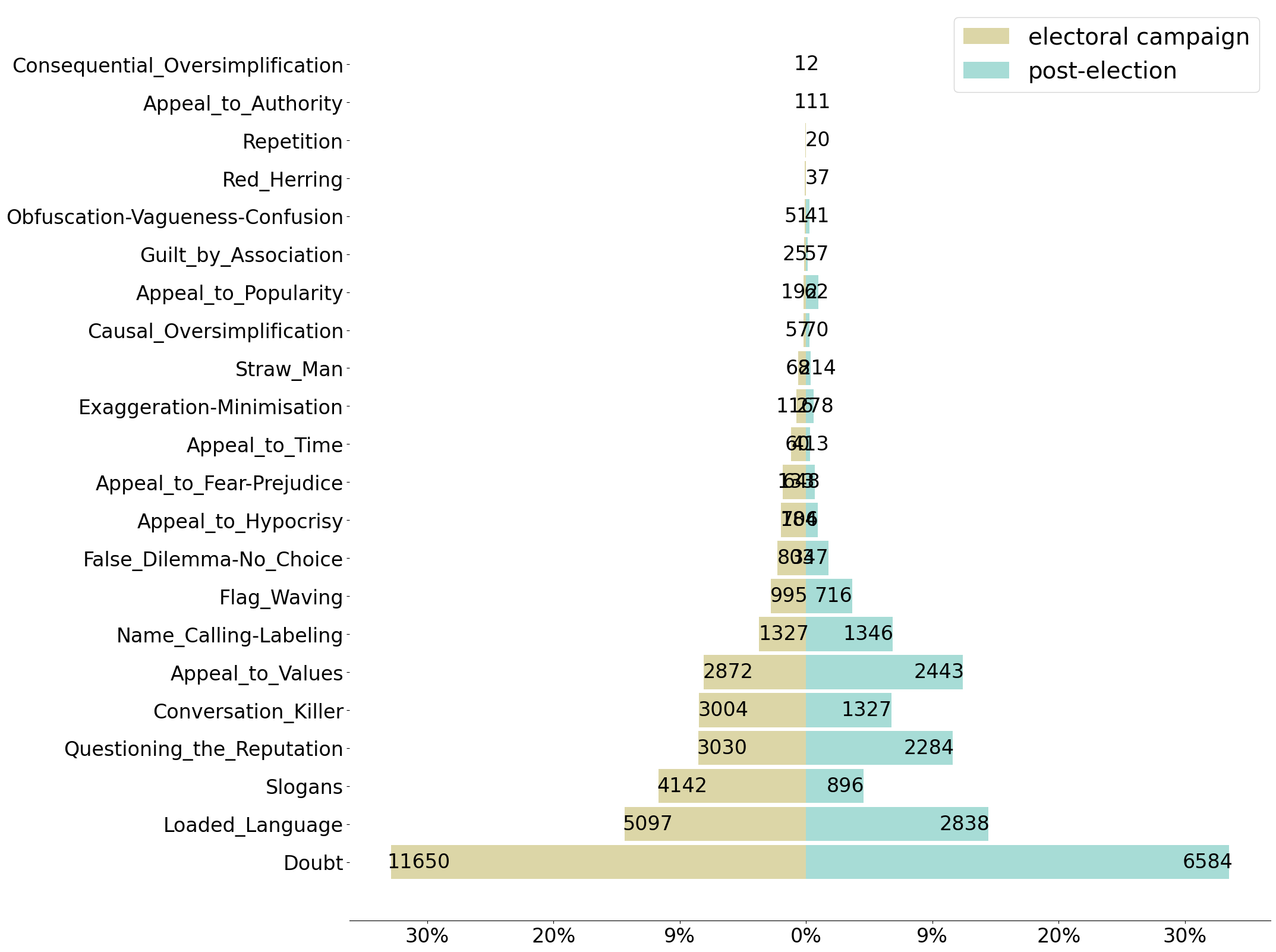}
            \caption{Vulnerability to propaganda techniques during the electoral campaign and post election for spurious swing voters.}
            \label{fig:tech_da_noreal}
        \end{figure}
\end{itemize}

\noindent Taking into account the existence of apolitical communities and recognizing that their members can also be vulnerable to political propaganda techniques despite not being part of a political community, we also analyzed the shifts from apolitical to political communities and vice versa.

\begin{itemize}
    \item In the case of shifts from \textit{apolitical to political} communities, we noticed that the use of \textit{appeal to values} increased consistently when transitioning to a political community in the post-election period, while the use of \textit{conversation killer} and \textit{slogans} decreased (cf. Figure~\ref{fig:nantoparty_da}). This trend is also observed when focusing on an apolitical to political shift from the pre-campaign period to the post-election period (cf. Figure~\ref{fig:nantoparty_ba}).
    The most common community shift in this context is towards (L, FdI), both between the campaign and the post-election periods and between the pre-campaign and the post-election periods.

    \begin{figure}[t!]
        \centering

        \begin{subfigure}{.45\columnwidth}
            \includegraphics[width=\linewidth]{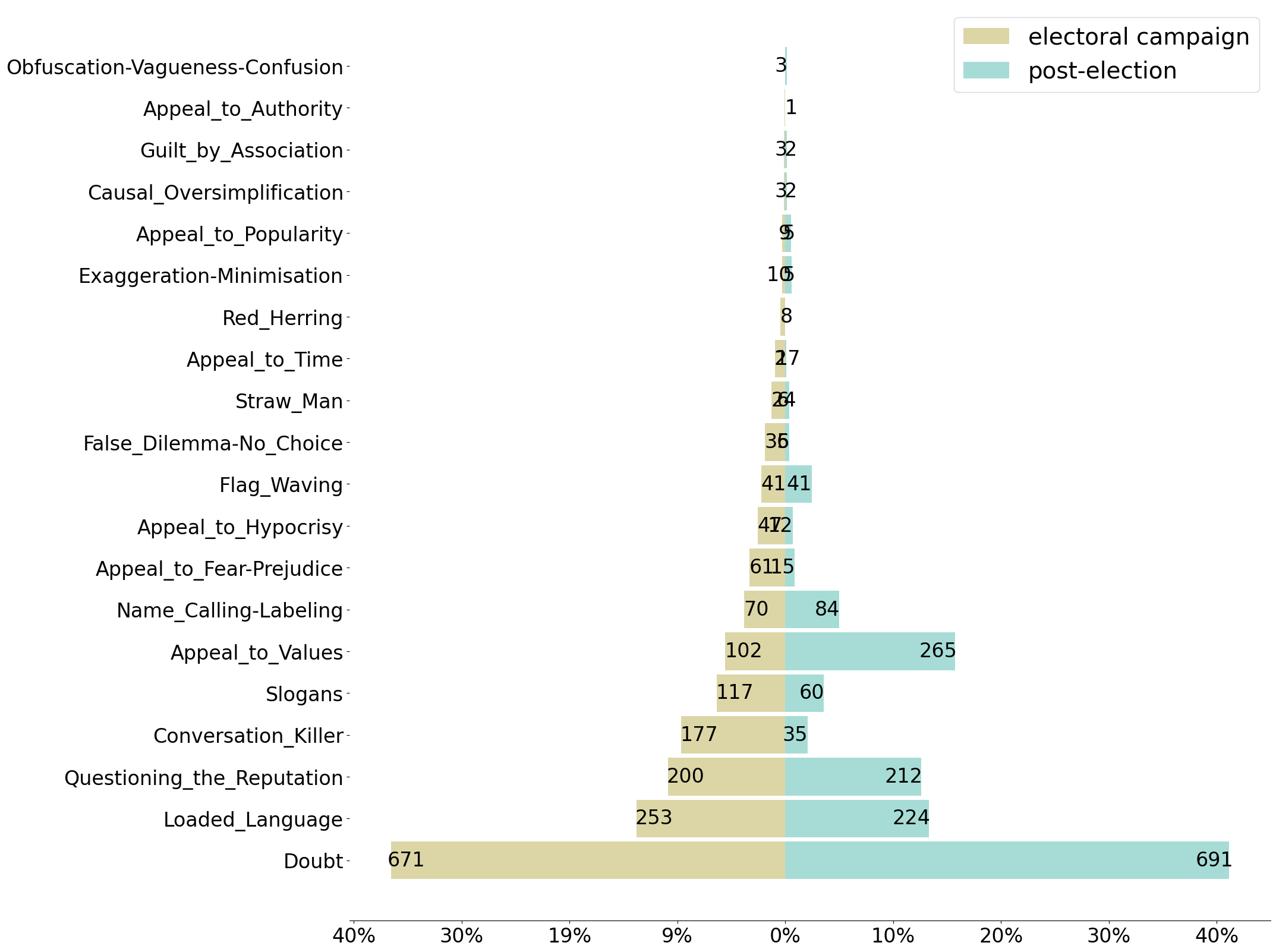}
            \caption{Electoral campaign vs. post election.}
            \label{fig:nantoparty_da}
        \end{subfigure}
        \begin{subfigure}{.45\columnwidth}
            \includegraphics[width=\linewidth]{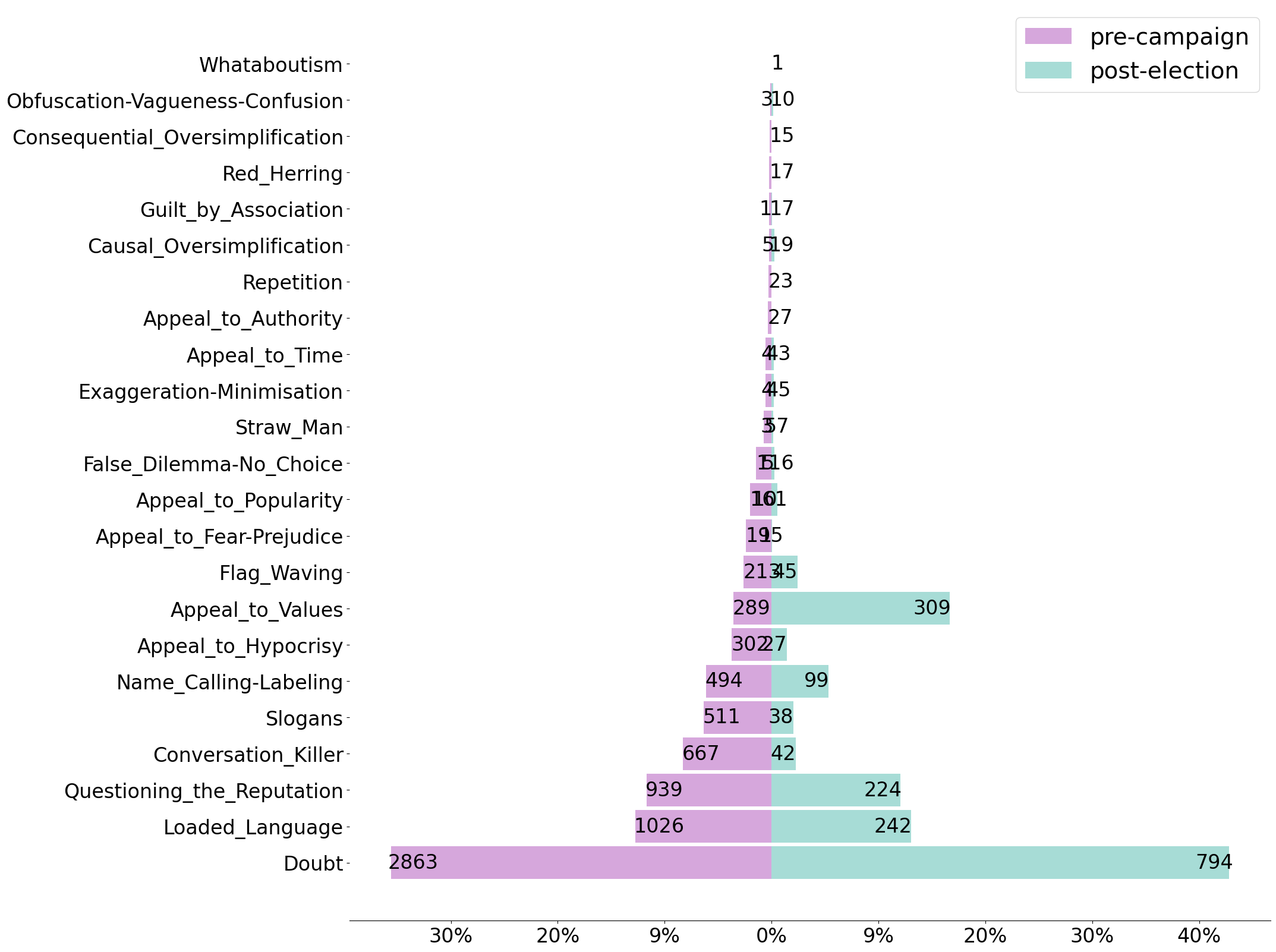}
            \caption{Pre-campaign vs. post election.}
            \label{fig:nantoparty_ba}
        \end{subfigure}

        \caption{Vulnerability to propaganda techniques during the electoral campaign vs. post-election and pre-campaign vs. post-election for apolitical-to-political swing voters.}
        \label{fig:nantoparty_both}
    \end{figure}

    \smallskip

    \item When focusing on the shift from \textit{political to apolitical} communities, we noticed a reduction in vulnerability to \textit{Questioning the reputation} from the campaign to post-election phases (cf. Figure~\ref{fig:partytonan_da}), and a consistent increase in vulnerability to \textit{Slogans} when shifting from the pre-campaign to post-election periods (cf. Figure~\ref{fig:partytonan_ba}). Most of the political shifts, both from the campaign to post-election and from pre-campaign to post-election, originate from the (PD, AVS) community.

    \begin{figure}[h!]
        \centering

        \begin{subfigure}{.45\columnwidth}
            \includegraphics[width=\linewidth]{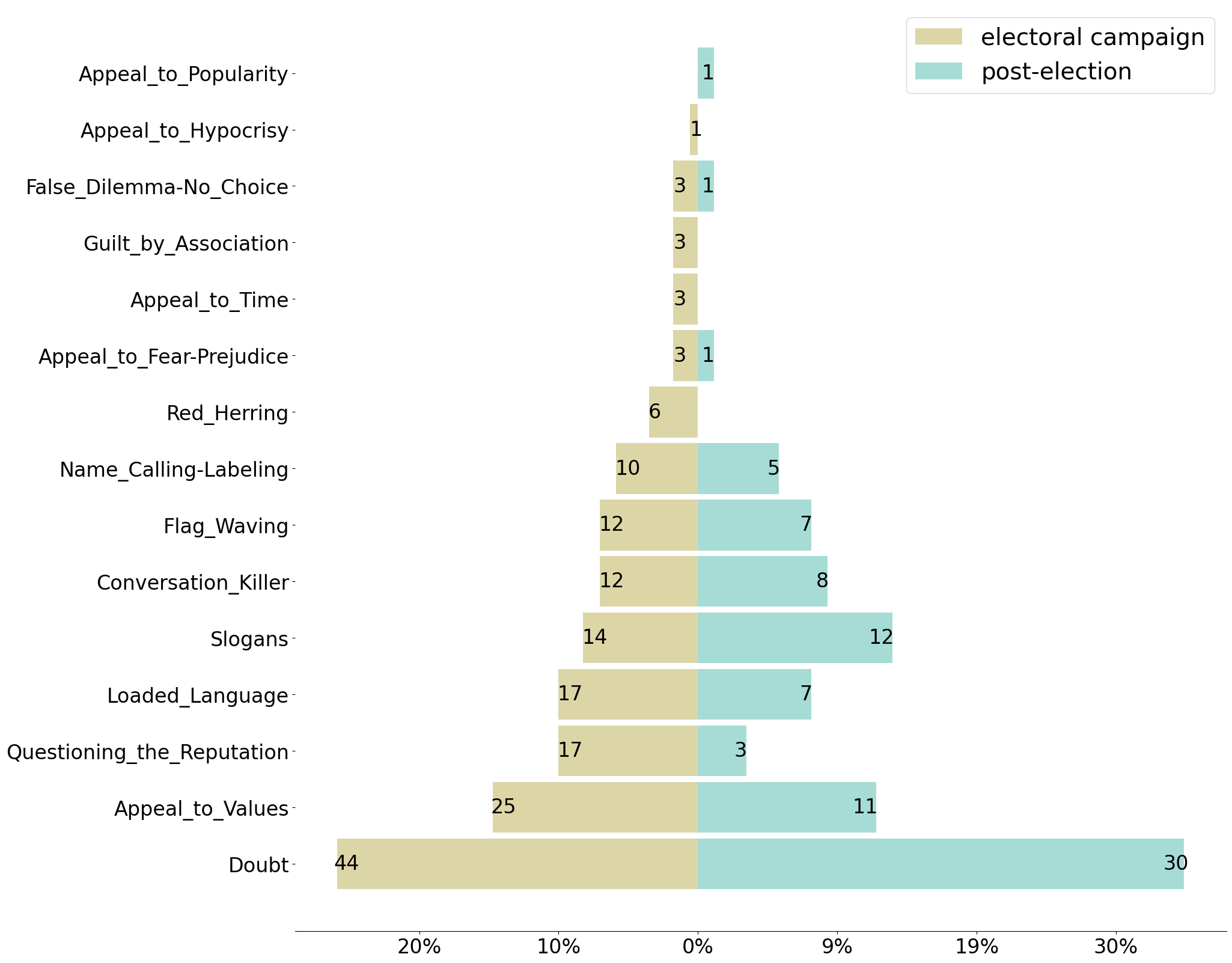}
            \caption{Campaign vs. post-elections.}
            \label{fig:partytonan_da}
        \end{subfigure}
        \begin{subfigure}{.45\columnwidth}
            \includegraphics[width=\linewidth]{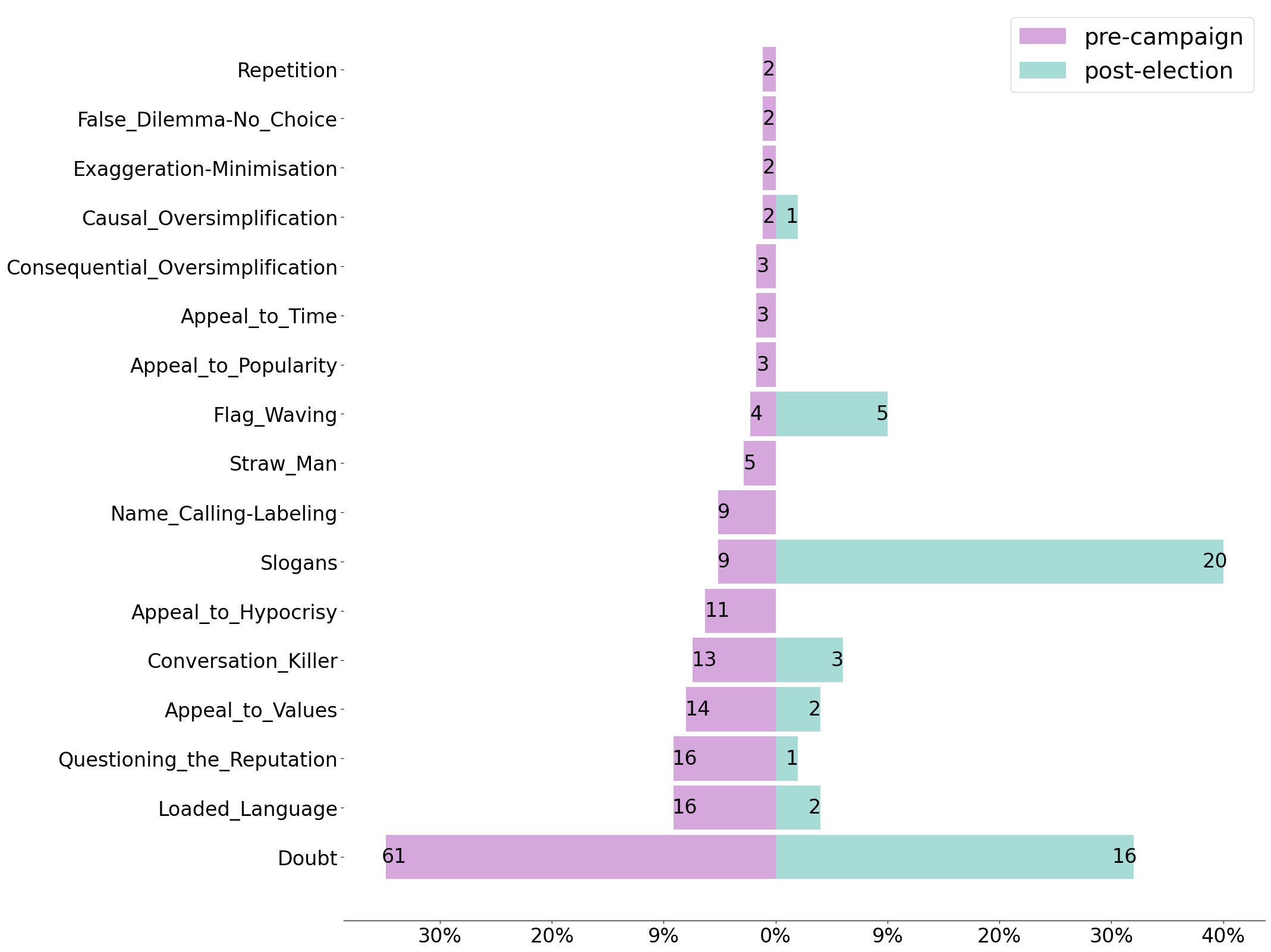}
            \caption{Pre-campaign vs. post-elections.}
            \label{fig:partytonan_ba}
        \end{subfigure}

        \caption{Vulnerability to propaganda techniques during the campaign vs. post-elections and pre-campaign vs. post-elections for political-to-apolitical swing voters.}
        \label{fig:partytonan_both}
    \end{figure}

\end{itemize}

\newpage
\section{Discussion} 
\label{sec:discussion}
In this section, we will refer to the RQs defined in the Introduction and elaborate on them based on our findings.

\medskip

\noindent \textit{\textbf{RQ$_1$: The retweet networks built on top of the 2022 Italian elections Twitter discourse show a highly skewed in-degree distribution, zero reciprocity, and negative degree assortativity, indicating that influential users, such as politicians, are predominantly retweeted by less popular ones and do not interact with each other. These networks also exhibit low density, transitivity, and clustering values, yet remain well-connected.
The community composition is extremely homogeneous in terms of political affiliations, and communities' evolution reveals a generally stable political landscape.
}}
A well-known phenomenon on social media platforms is that a small fraction of users generate most of the original content~\cite{Nielsen_pyramid}. On Twitter, for instance, most users tend to engage more in retweeting or liking posts rather than creating their own~\cite{Antelmi_TheWebConf_2019}. These dynamics result in a few prolific users generating the majority of original posts while the rest of the user base amplifies these messages through retweets or likes. 
We found that interactions were predominantly one-directional in our dataset. Users generally retweeted content from well-known users, especially political representatives, rather than engaging in reciprocal communication. This phenomenon suggests that the flow of information was largely from political figures to the general public.
As discussed in Section~\ref{subsec:network_results}, These dynamics are evident in the highly skewed in-degree distribution, reciprocity values of zero (indicating that mutual connections, where user A retweets user B and vice versa, are virtually nonexistent), and negative degree assortativity values, meaning that users with high degrees (frequently retweeted users) are often connected to users with low degrees (those who rarely receive retweets) and that these popular users do not interact extensively among themselves. 
All backbones exhibit low density, transitivity, and clustering coefficient values. In particular, low transitivity and clustering coefficients suggest that the networks lack tightly-knit groups, reinforcing the idea of a few central nodes dominating the network. Despite these low values, all networks are quite connected.

\smallskip

All three retweet networks exhibit highly clustered structures, highlighted by consistently robust modularity and coverage metrics, along with negligible conductance values. Interestingly, but not surprisingly, the composition of the political communities is strongly homogeneous in terms of political affiliation.

Specifically, during the pre-campaign period, each community almost perfectly corresponds to a specific political party, indicating a clear separation of users based on their political leanings. During the electoral campaign, we observe the emergence of a new right-wing coalition, reflecting the changing political landscape, such as parties forming alliances. In the post-election period, a more fragmented situation arises, where nearly every community can once again be associated with a single party, suggesting a return to traditional party lines.
Despite these shifts, the overall dynamics of political communities remained relatively stable over time as most users either stayed within their original communities or moved to other communities that aligned with their existing political affiliations. This stability indicates that while there are changes in political alignments during key periods, the core political affiliations of users remain largely consistent.

\medskip

\noindent \textit{\textbf{RQ$_2$: The popularity of politicians is dynamically defined during election times. The use of propaganda techniques increases in the pre-election campaign and campaign periods and tends to peak just before the election day.}}
Section \ref{subsubsec:centrality} has explored the centrality of political representatives within their social network in terms of engagement and relevance throughout the election periods. We observed an increase in engagement during the election campaign and a significant decrease after the elections, underlining some unexplored potential for higher post-election engagement. However, a limited set of politicians maintains good levels of audience engagement across the three considered election periods. Moving the focus to relevance, we observed that most political representatives maintained their pre-campaign role during the election campaign phase, while shifts of relevance happened after the elections and when comparing the pre-campaign to the post-election periods.
Politicians exhibited varying degrees of local importance, with only a selected few occupying the ``core" of the network and the majority remaining in the periphery due to their lower local significance. This duality between shifting peripheries and persistent cores was particularly notable: continuous changes in core-periphery distributions and composition throughout the elections highlighted a particularly dynamic political scenario from a social debate perspective. Meanwhile, certain politicians consistently maintained their positions at the network's core, indicating greater resilience to shifts in relevance dynamics and a more consolidated influence.

\smallskip

Having observed some dynamic popularity trends of politicians throughout the election period, we focused on the language they used in their social media discourse in terms of propaganda techniques. Overall, we noticed that the volume of tweets containing propaganda techniques increases during the pre-election campaign and the campaign phases, then peaks just before the election day and decreases after the elections. Doubt is the most commonly used propaganda technique, followed by loaded language, appeal to values, and slogans. Different political wings tend to use different propaganda techniques more than others: right-wing parties' representatives use less doubt and more flag-waiving and slogans than the baseline in their online communication, while center-left parties' representatives tend to use doubt more commonly and flag-waiving less frequently compared to the baseline.

To understand whether the change in popularity of politicians can be related to their use of propaganda techniques throughout the elections, we focused on those experiencing a significant variation of popularity during the election times. While it is not possible to establish a clear association pattern between the use of specific propaganda techniques and variation in centrality, we observed that politicians who observed a variation in popularity also changed their communication tone, manifested as a diverse set of propaganda techniques employed.

\medskip

\noindent \textbf{\textit{RQ$_3$: A notable portion of users shifted their political views, and there is evidence suggesting that these users are more vulnerable to propaganda tweets than non-swing voters.}}
As discussed in Section~\ref{subsubsec:swing_voters_detection_results}, we observed a significant percentage of hard swing voters who migrated from right-wing to left-wing affiliations and vice-versa. Specifically, these users represent 7.62\% of active users evaluated in the periods before and during the electoral campaign and 6.43\% of users who remained active from during the campaign to after the elections. 
When examining all transitions regardless of the specific periods, we found that shifts from left to right-wing affiliations involved both PD and AVS. Specifically, many users transitioned from PD and [AVS, PD] to [Az-Iv, FdI] and from [AVS, PD] to Az-Iv and [M5s, FdI]. Conversely, the most common transitions from right to left affiliations occurred from Az-Iv, [L, Az-Iv], [M5s, FdI], and FdI to [AVS, PD], and from FdI to PD.
Additionally, we noted a substantial presence of users transitioning to political communities in the post-election period. Most of these users moved to a right-wing coalition, which won the elections, indicating a significant shift in political alignment following the election results.
Overall, these results showed that although the general political landscape of the Italian elections remained relatively stable on the Twitter social platform, a significant portion of active users experienced a shift in their political views during these key periods. This outcome suggests a notable degree of political fluidity among a subset of the electorate despite the broader stability.

\smallskip

In terms of propaganda vulnerability, swing voters tended to retweet propaganda tweets more than non-swing voters, particularly just before their political opinion shift took place. This increased retweeting activity indicates a higher susceptibility to propaganda during periods of political transition. Notably, different categories of swing voters exhibited increased vulnerability to specific propaganda techniques.
Specifically, we found that \textit{hard swing voters} were more vulnerable to loaded language, slogans, and name-calling/labeling post-elections compared to the pre-election campaign. This shift indicates that post-election, these users are more influenced by emotionally charged and derogatory rhetoric.
\textit{Soft swing voters}, who display milder shifts in political views, are noticeably more vulnerable to certain propaganda techniques after the elections. Compared to the pre-election period, they are more likely to be influenced by conversation-killer tactics, which stifle debate, and appeals to values, which leverage moral or ethical arguments.
\textit{Spurious swing voters}, whose political shifts remain closely aligned with the original party or coalition, showed an increased susceptibility to the appeal to values technique after the election compared to during the campaign. This suggests that moral and ethical arguments become more persuasive to these voters in the post-election period.
Users transitioning from \textit{apolitical to political} communities demonstrated a higher vulnerability to appeal to values propaganda after the elections. As these users become politically engaged, they are more likely to be influenced by messages that align with their moral or ethical beliefs, more so than during the pre-campaign and campaign periods.
Conversely, users who moved from \textit{political to apolitical} communities showed an increased susceptibility to slogans post-election. This suggests that even as they disengage from active political communities, they remain influenced by simple, catchy phrases that encapsulate political messages, perhaps because these slogans are pervasive and easily digestible.
These observations highlight the varying degrees and types of propaganda vulnerability among different categories of swing voters, emphasizing the nuanced ways in which political engagement and susceptibility to propaganda evolve over time and across different electoral phases.

\section{Conclusion} 
\label{sec:conclusions}
This study examined the Twitter discourse during the 2022 Italian general elections, revealing a predominantly stable political landscape with a minority of users demonstrating significant shifts in political preferences. These users, identified as swing voters, are more susceptible to propaganda messages spread by politicians than those who maintain consistent voting preferences. Our findings indicate that politicians vary their use of propaganda techniques throughout the election period, changing their online popularity at the same time.

To our knowledge, this is the first analysis of the online behavior of swing voters in terms of susceptibility to propaganda techniques within the Italian political context. Despite the novelty of our contribution, we acknowledge several limitations. First, our dataset was collected using a set of hashtags that may have excluded some relevant information. This dataset reflects only the online presence and interactions of voters with politicians, omitting real-life political participation and expressed preferences during the elections. Second, we did not account for information sources other than politicians, labeling communities that did not involve such users as non-political. A more comprehensive perspective would include an exploration of news sources and other relevant accounts and their roles in political discourse.

Future research will delve deeper into the motivations behind voters' preference shifts, considering factors such as personality traits, language use, and socio-economic elements. Additionally, we plan to focus on changes in parties' political manifestos throughout the election period and investigate whether swing voters tend to follow parties with similar agendas. Furthermore, we aim to integrate this study into the broader computational politics field to achieve a more robust validation of our findings.

\medskip

\noindent
The experimental code developed with this work is available at \href{https://github.com/ariannap13/hidden-swing-voters}{https://github.com/ariannap13/hidden-swing-voters}.

\clearpage

\bibliography{sn-bibliography}

\end{document}